\documentclass[prd,aps,floatfix,notitlepage,reprint,superscriptaddress,twocolumn,nofootinbib,preprintnumbers]{revtex4-2}
\pdfoutput=1
\usepackage{graphicx,bm,graphics,amsmath,bm,natbib,multirow,hyperref,float,fancyhdr,amssymb,amsfonts}
\usepackage[usenames,dvipsnames]{color}
\usepackage[caption = false]{subfig}
\hypersetup{colorlinks=true,citecolor=blue,linkcolor=blue}

\newcommand{\be}{\begin{equation}}
\newcommand{\ee}{\end{equation}}

\newcommand{\hMpc}{\,h\,{\rm Mpc}^{-1}}
\newcommand{\Mpch}{\,h^{-1}{\rm Mpc}}
\newcommand{\vev}[1]{\langle #1 \rangle}
\newcommand{\shifted}[1]{\tilde{#1}}
\newcommand{\Perr}{P_{\rm err}}

\def\k{{\boldsymbol{k}}}

\def\q{{\boldsymbol{q}}}
\def\p{{\boldsymbol{p}}}

\def\z{{\boldsymbol{z}}}

\def\n{{\boldsymbol{n}}}
\def\vpsi{{\boldsymbol{\psi}}}

\def\G{{\mathcal{G}}}

\begin{document}

\preprint{CERN-TH-2022-127}

\title{Modeling HI at the field level}
\author{Andrej Obuljen}\email{andrej.obuljen@uzh.ch}
\affiliation{Institute for Computational Science, University of Zurich, Winterthurerstrasse 190, 8057 Zurich, Switzerland}
\author{Marko Simonovi{\'c}}\affiliation{Theoretical Physics Department, CERN, 1 Esplanade des Particules, Geneva 23, CH-1211, Switzerland}
\author{Aurel Schneider}
\author{Robert Feldmann}
\affiliation{Institute for Computational Science, University of Zurich, Winterthurerstrasse 190, 8057 Zurich, Switzerland}
\date{\today}

\begin{abstract}
We use an analytical forward model based on perturbation theory to predict the neutral hydrogen (HI) overdensity maps at low redshifts. We investigate its performance by comparing it directly at the field level to the simulated HI from the IllustrisTNG magneto-hydrodynamical simulation TNG300-1 ($L=205\Mpch$), in both real and redshift space. We demonstrate that HI is a biased tracer of the underlying matter field and find that the cubic bias model describes the simulated HI power spectrum to within 1\% up to $k=0.4 \;(0.3) \hMpc$ in real (redshift) space at redshifts $z=0,1$. Looking at counts in cells, we find an excellent agreement between the theory and simulations for cells as small as 5 $h^{-1}$ Mpc. These results are in line with expectations from perturbation theory and they imply that a perturbative description of the HI field is sufficiently accurate given the characteristics of upcoming 21cm intensity mapping surveys. Additionally, we study the statistical properties of the model error – the difference between the truth and the model. We show that on large scales this error is nearly Gaussian and that it has a flat power spectrum, with amplitude significantly lower than the standard noise inferred from the HI power spectrum. We explain the origin of this discrepancy, discuss its implications for the HI power spectrum Fisher matrix forecasts and argue that it motivates the HI field-level cosmological inference. 
On small scales in redshift space we use the difference between the model and the truth as a proxy for the Fingers-of-God effect. This allows us to estimate the nonlinear velocity dispersion of HI and show that it is smaller than for the typical spectroscopic galaxy samples at the same redshift. 
Finally, we provide a simple prescription based on the perturbative forward model which can be used to efficiently generate accurate HI mock data, in real and redshift space. 
\end{abstract}

\maketitle

\section{Introduction}
The main goal of current galaxy surveys is to measure cosmological parameters using galaxy clustering as a probe of the underlying matter density field over large volumes and large redshift ranges. Spectroscopic galaxy redshift surveys such as DESI~\cite{DESI} and Euclid~\cite{Euclid} will measure approximately $30$ million galaxy positions, an order-of-magnitude better than the largest currently available spectroscopic sample of BOSS~\cite{BOSS, eBOSS}. 

Another way to probe the distribution of matter is to use neutral hydrogen (HI), relying on the technique of 21cm intensity mapping (IM) \cite{im1_Bharadwaj,im2_Bharadwaj,im3_chang,im4_peterson,Bull}. In this approach the aim is to collect an integrated HI signal from the hyperfine spin-flip transition of atomic hydrogen at $21$cm across different redshifts. A number of upcoming 21cm surveys such as CHIME~\cite{CHIME}, HIRAX~\cite{HIRAX}, SKA~\cite{SKACosmo}, Tianlai~\cite{Tianlai} and proposed PUMA~\cite{PUMA} will use the 21cm line to probe the large-scale clustering of HI across the wide range of redshifts in the post-reionization ($z<6$) universe, hopefully providing some of the tightest constraints on the $\Lambda$CDM cosmological model and its extensions.

One of the keys in achieving these ambitious goals is the simple, reliable and accurate theoretical description of the nonlinear density field of HI in terms of the initial conditions.
The clustering of HI is somewhat different with respect to the standard galaxy clustering due to the particular form of the HI-halo connection. On average, the total HI mass in halos increases with the halo mass as a power law with an exponential cutoff at small halo masses. Together with the shape of the halo mass function, this results in HI signal originating mainly from small and intermediate halo masses~$M_h\sim 10^{10-12}\ [h^{-1} M_\odot]$. This has two important consequences. First, as these halos are very abundant, the expected HI shot-noise is low~\cite{CastorinaPaco,PacoTNG}. Second, the fraction of HI in central versus satellites significantly depends on the halo mass~\cite{PacoTNG,Modi}. For this reason it is hard to estimate whether the Fingers-of-God (FoG) effects in HI are larger or smaller compared to the FoG in typical galaxy samples, even though the velocity dispersion of HI is smaller than that one of the matter~\cite{PacoTNG}. Finally, it is worth noting that, in contrast to the discrete nature of galaxy catalogs, the observed HI is smoothed by the angular and/or frequency resolution, and therefore represented by a continuous field.

However, despite these peculiarities, the HI is still expected to be a biased tracer of the underlying matter field. Since the nonlinear evolution of HI is expected to be local in space~\cite{Modi_Hidden_Valley} and to obey the Equivalence Principle, the same bias expansion formalism developed for galaxies (see~\cite{Bias_review} for a review) is valid for the HI as well. This approach is inherently perturbative. Therefore, one can use analytical techniques to find a relatively simple relation between initial conditions and the final HI field, relying only on the basic features of gravitational dynamics and symmetries of the system. The major advantage of perturbation theory is its universality and robustness. It is valid for {\em any} (local) baryonic physics, up to only a handful of free nuisance parameters which capture all the intricacies of the galaxy formation process, as formalized in the effective field theory approach to galaxy clustering~\cite{Baumann:2010tm,Carrasco:2012cv,Senatore:2014via,Senatore:2014eva,Foreman:2015lca,Perko:2016puo,Cabass:2022avo}. 

Nevertheless, it is useful to carefully test these predictions against reality, since the relevant scales that determine the range of validity of perturbation theory can depend on the type of tracer considered. The main goal of this paper is to provide such careful analysis for the case of HI at low redshifts and in the context of planned IM surveys. To that end, we use the perturbative forward modelling to predict the HI maps and compare it to the state-of-the art IllustrisTNG magneto-hydrodynamical simulations which have been instrumental in studying the clustering of HI in the late universe~\cite{TNGa,TNGb,TNGc,TNGd,TNGe}. Such field level comparison has a number of advantages over the standard fit to the power spectrum or other $n$-point functions. Firstly, choosing the same initial conditions for theory and simulations, one does not pay the price of cosmic variance. Secondly, in order to match the two realizations, amplitudes and phases of all Fourier modes have to be correct. This is a much more stringent test than looking at a particular summary statistics which unavoidably involves averaging over Fourier modes. Finally, 
at the field level one can also study observables such as counts in cells, which are not simply related to the $n$-point functions. 
For all these reasons, the field level methods have been exploited to study the performance of perturbation theory for dark matter~\cite{Baldauf:2015tla,Baldauf:2015zga} as well as biased tracers, in real and redshift space~\cite{Schmittfull,Schmittfull_RSD}. In the context of HI, this approach has been already used in several works~\cite{Modi,McQuinn,Thesan}, mainly at higher redshifts ($z>2$). While based on the same ideas as in these earlier papers, our implementation of the field level analysis is slightly different, and we will comment on these differences in what follows. Other approaches, based on machine learning techniques, have also been used to predict HI at the field level \cite{wadekar, ember}.

Another important advantage of the field level comparison is that it provides a clear path to study the difference between the theory and simulations. On large scales, where the unaccounted nonlinearities are very small, this difference is a true stochastic noise of galaxy formation, which depends on the number density and the type of tracers being observed. On small scales, this difference is dominated by the nonlinear evolution and can be used as a proxy of the truly nonlinear effects that cannot be captured by perturbation theory. We will show how both of these regimes can be exploited in order to gain a deeper understanding of clustering properties of HI. Those include measuring the true amplitude of the shot noise and estimating the HI nonlinear velocity dispersion.

Finally, perturbation theory can provide a good starting point to describe the HI density field even in the nonlinear regime. This is achieved by projecting the HI map to a few perturbative templates, allowing the coefficients in this projection, the so-called transfer functions, to be free functions of scale. These transfer functions are smooth and can be fitted by low-degree polynomials. As long as this simple expansion provides a decent phenomenological description of the nonlinear field, the transfer functions can be used to generate realistic HI mock data. We exploit this fact and provide a publicly available code \texttt{Hi-Fi mocks}\footnote{\url{https://github.com/andrejobuljen/Hi-Fi_mocks}} that can be used by the community to efficiently generate maps of HI whose properties resemble the one from the IllustrisTNG simulations, but with arbitrary volumes, different realizations of the initial conditions or even slightly different cosmologies.

The outline of the paper is the following. In~\S\ref{sec:theory} we describe the theoretical model that we use to make HI maps given some cosmology and a realization of the initial conditions. In~\S\ref{sec:sims} we describe the IllustrisTNG simulations that we use as a reference point of comparison. The results in real and redshift space for the best fit realization, the power spectrum analysis and counts in cells are presented in~\S\ref{sec:results}. In~\S\ref{sec:HInoise} we discuss in detail the HI noise properties, including the amplitude of the noise on large scales and nonlinear velocity dispersion. In~\S\ref{sec:mocks} we show how to create fast and accurate HI mock data based on perturbative templates and in~\S\ref{sec:conclusions} we conclude and lay out directions for future research.

\section{Theoretical model}
\label{sec:theory}
In this section we briefly describe the perturbative model that we use to predict HI maps at a given redshift, for a given cosmology and initial conditions. We use the exact same formalism of~\cite{Schmittfull} in real space and of~\cite{Schmittfull_RSD} in redshift space and we refer the reader to these papers for more details. Before we provide the main ingredients of this approach, let us point out that it is designed to reproduce the correct one-loop power spectrum and the tree-level bispectrum, including all relevant dark matter nonlinearities, effective field theory counterterms, bias parameters and infrared resummation. Therefore, the one-loop power spectrum of this perturbative forward model is identical to the one computed by the nonlinear extensions of the Boltzmann codes, such as \texttt{CLASS-PT}~\cite{Chudaykin:2020aoj}, \texttt{PyBird}~\cite{DAmico:2020kxu} or \texttt{velocileptors}~\cite{Chen:2020fxs}.

The main difference in making a perturbative prediction at the field level compared to the $n$-point functions is the treatment of large displacements. While by the Equivalence Principle the large displacements do not affect the smooth part of the correlation functions~\cite{Peloso:2013zw,Kehagias:2013yd,Creminelli:2013mca,Creminelli:2013poa} and impact only the features such as BAO~\cite{Senatore:2014via,Baldauf:2015xfa,Vlah:2015zda,Blas:2016sfa}, their effect at the field level is much more dramatic. Inspired by the Lagrangian perturbation theory which treats the large displacements nonperturbatively, in this paper we use the shifted operators of~\cite{Schmittfull,Schmittfull_RSD} as the building blocks for describing the HI realizations. These shifted operators, which we denote with the tilde signs, have the following form (in redshift space) 
\be
\label{eq:shifted_def}
\tilde{\mathcal{O}}(\bm{k},\hat{\bm{n}}) \equiv \int d^3\bm{q}\; \mathcal{O}(\bm{q}) e^{-i\bm{k}\cdot(\bm{q}+\bm{\psi}_1(\bm{q}) + f \hat{\bm{n}} (\bm{\psi}_1(\bm{q}) \cdot \hat{\bm{n}}) ) },
\ee
where $q$ is the Lagrangian coordinate in the initial conditions, $\bm{\psi}_1$ is the Zel'dovich displacement field, $\hat{\bm{n}}$ is the line of sight direction and $f$ is the logarithmic growth function. These operators are the building blocks in  calculating Eulerian density field using Lagrangian perturbation theory~\cite{2008PhRvD..78h3519M,clpt,2016JCAP...12..007V,2021JCAP...03..100C}. Their statistics can be calculated straightforwardly~\cite{Chen:2020fxs}, and the results can be shown to be equivalent to the IR-resummed Eulerian counterparts~\cite{Schmittfull,Chen:2020fxs}. Note that for simplicity we suppress the explicit time dependence, but all quantities in this expression have to be evaluated as functions of time. For instance, in order to obtain the Zel'dovich displacement field we use the initial density field $\delta_1$ rescaled linearly to a given output redshift ($z=0,1$ in our case):
\be
\bm{\psi}_1(\bm{k},z)=\frac{i\bm{k}}{k^2}\delta_1(\bm{k},z).
\ee
Clearly, in order to obtain the shifted operators in real space, one can set $f=0$ in Eq.~\eqref{eq:shifted_def} and drop the dependence on the line of sight. 

Let us now turn to the question of which operators do we use in our forward model. We focus on the leading orders in perturbation theory, needed to ensure that the one-loop power spectrum and the tree-level bispectrum are predicted correctly. This implies that we have to keep all relevant terms up to cubic order in small density fluctuations. However, there is a simple trick which allows us to absorb the contribution of the cubic operators into the scale-dependent coefficients of the shifted linear field. To see this, let us decompose a generic cubic operator in the following way 
\be
\tilde{\mathcal O}_3 = \frac{\langle \tilde{\mathcal O}_3  \tilde{\delta}_1 \rangle}{\langle \tilde{\delta}_1 \tilde{\delta}_1 \rangle} \tilde{\delta}_1 + \left( \tilde{\mathcal O}_3 - \frac{\langle \tilde{\mathcal O}_3  \tilde{\delta}_1 \rangle}{\langle \tilde{\delta}_1 \tilde{\delta}_1 \rangle} \tilde{\delta}_1 \right) .
\ee
We can define the second term in the brackets to be orthogonal to $\tilde{\delta}_1$
\be
\tilde{\mathcal O}_3^\perp \equiv \tilde{\mathcal O}_3 - \frac{\langle \tilde{\mathcal O}_3  \tilde{\delta}_1 \rangle}{\langle \tilde{\delta}_1 \tilde{\delta}_1 \rangle} \tilde{\delta}_1 \;,
\ee
since, by definition, $\langle \tilde{\mathcal O}_3^\perp \tilde{\delta}_1 \rangle = 0$. This means that these orthogonal contributions do not contribute to the one-loop power spectrum and can be neglected at this order in perturbation theory. The only impact of cubic operators is through the scale-dependent coefficient multiplying $\tilde{\delta}_1$, which we refer to as the transfer functions
\be
\beta_3(\bm k,\bm{\hat n}) \equiv \frac{\langle \tilde{\mathcal O}_3  \tilde{\delta}_1 \rangle}{\langle \tilde{\delta}_1 \tilde{\delta}_1 \rangle} \;.
\ee
These functions can be computed analytically, without the need to perform the forward modeling for cubic operators on the grid. In conclusion, provided that the proper transfer functions are used, the only operators that have to be generated at the field level are $\mathcal{O}=\{1,\delta_1, \delta_2\equiv(\delta_1^2-\sigma_1^2),\mathcal{G}_2\}$, where $\sigma_1^2\equiv\langle\delta_1^2 \rangle$ is the r.m.s.~fluctuation of the linear density field, while
\be
\mathcal{G}_2\equiv\left[\frac{\partial_i \partial_j}{\nabla^2} \delta_1\right]^2 - \delta_1^2
\ee
is the second order bias operator related to the tidal field. In some analyses we will also explicitly include $\mathcal{O}=\delta_1^3$. While this operator is not needed for the model for the reasons we have just explained, it is the only cubic term relevant for studying the properties of the noise, given that it has a flat auto-spectrum on large scales. 

In order to simplify calculations in practice, it is convenient to use the basis of operators that are all orthogonal to each other, such that $\langle \tilde{\mathcal O}_a^\perp \tilde{\mathcal O}_b^\perp \rangle = 0$, for any $a$ and $b$. This is achieved by a simple linear transformation, which does not change the statistical properties of the map. We will use such orthogonal basis in all our analyses. 

\subsection{Real-space model}
\label{sec:real_space_model}
Let us focus on real space first. The full HI field can be written as a sum of two contributions
\be
\delta_\mathrm{HI}(\bm{k}) = \delta_\mathrm{HI}^{\rm model}(\bm{k}) + \epsilon_\mathrm{HI}(\bm{k}) \;, 
\ee
where~$\delta_\mathrm{HI}^{\rm model}$ is the theoretical model and~$\epsilon_\mathrm{HI}$ is the model error which we assume to be uncorrelated with~$\delta_\mathrm{HI}^{\rm model}$. For example, on large scales, we expect~$\epsilon_\mathrm{HI}$ to be dominated by the stochastic noise due to discreteness of galaxies as a tracer. We follow ref.~\cite{Schmittfull} and use the following model for the HI field in terms of shifted and orthogonalized operators $\tilde{\mathcal{O}}^\perp(\bm{k})$:
\be
\label{eq:real_model}
\begin{split}
\delta_\mathrm{HI}^{\rm model}(\bm{k}) = \beta_1(k) & \tilde{\delta}_1(\bm{k}) + \beta_2(k)\tilde{\delta}_2^\perp(\bm{k}) \\ & + 
\beta_{\mathcal{G}_2}(k)\tilde{\mathcal{G}}_2^\perp(\bm{k}) +\beta_3(k)\tilde{\delta}_3^\perp(\bm{k}),
\end{split}
\ee
where $\beta_i(k)$ are transfer functions. At next-to-leading order in perturbation theory these functions have the following form~\cite{Schmittfull} (see also Appendix~\ref{app:real_app})
\be
\begin{split}
\label{eq:beta1_model}
\beta_1(k) = b_1 + &\ c_s^2 k^2 + b_2 \frac{\vev{\shifted\delta_1 \shifted \delta_2}}{\langle\shifted \delta_1 \shifted \delta_1\rangle}+ b_{\mathcal{G}_2} \frac{\vev{\shifted\delta_1 \shifted {\mathcal{G}}_2}}{\vev{\shifted\delta_1 \shifted\delta_1 }} \\ + &\ b_{\Gamma_3} \frac{\vev{\shifted\delta_1 \shifted {\Gamma}_3}}{\vev{\shifted\delta_1 \shifted\delta_1 }} - b_1 \frac{\vev{\shifted\delta_1 \shifted {\mathcal S}_3}}{\vev{\shifted\delta_1 \shifted\delta_1 }}, 
\end{split}
\ee
\be
\label{eq:beta_higher_model}
\beta_2(k) = b_2, \ \beta_{\mathcal{G}_2}(k) = b_{\mathcal{G}_2}, \ \beta_3(k) \ =\ b_3,
\ee
where $b_1, c_s^2, b_2, b_{\mathcal{G}_2}, b_{\Gamma_3}, b_3$ are constant free nuisance parameters. Note that the parameter $c_s^2$ is the sum of the dark matter one-loop counterterm and the so-called nonlocal bias. The shape of cubic operators $\Gamma_3$ and $\mathcal S_3$ can be found in Appendix~\ref{app:real_app}. Let us stress that~$\mathcal S_3$ comes from the second order displacement acting on the halo density field in Lagrangian coordinates, and therefore, by the Equivalence Principle, it is not multiplied by a new free parameter. 

We would like to emphasize once again that the exact form of the transfer function in Eq.~\eqref{eq:beta1_model} is needed in order to have the correct one-loop power spectrum. Sometimes in the literature the loop corrections are neglected and $\beta_1(k)$ is approximated as a constant~\cite{Modi}. While this may be a decent approximation at high redshifts ($z>2$), the loop corrections to $\beta_1(k)$ in the late universe can be significant. Furthermore, the $\tilde{\mathcal S}_3$ operator (usually neglected in the literature) is very important in order to get a reliable estimates of bias parameters, as an important additional cross check that our theoretical description makes sense. We will come back to this point in~\S\ref{sec:results} when we fit the transfer functions in real space using perturbative templates. 

Finally, let us comment on evaluating $\beta_1$ analytically. The IllustrisTNG simulation that we use as a reference point for comparison has the box size of $L=205\Mpch$. Given the periodic boundary conditions, this implies that the linear power spectrum has an infrared cutoff at $k_{\rm IR}=2\pi/L = 0.031 \; \hMpc$. In all theoretical calculations, the same cutoff has to be implemented. Note that this implies that the typical displacements are significantly smaller than in the real Universe. Cutting the power spectrum below $k_{\rm IR}$, we find that the velocity dispersion at redshift zero is $\sigma_v = 4.9 \; \Mpch$, roughly $20\%$ lower than $\sigma_v$ in $\Lambda$CDM and in excellent agreement with measurements from Illustris. Keeping this in mind, all correlators in Eq.~\eqref{eq:beta1_model} can be still evaluated following the standard methods for calculating the one-loop power spectrum in Lagrangian perturbation theory~\cite{Chen:2020fxs}, using the modified linear power spectrum with the appropriate infrared cutoff.

\subsection{Redshift-space model}
\label{sec:rsd_model_sec}
Let us next turn to redshift space. Following~\cite{Schmittfull_RSD}, we can write the perturbative model for the HI field as
\be
\begin{split}
\label{eq:RSD_model}
\delta_\mathrm{HI}^{s,{\rm model}} & (\bm{k,\hat n}) =  \delta_Z^s(\bm{k,\hat{n}}) - \frac{3}{7}f\tilde{\mathcal{G}}_2^\parallel(\bm{k,\hat{n}}) \\ + & \beta_1(k,\mu)\tilde{\delta}_1(\bm{k,\hat{n}}) + \beta_2(k,\mu)\tilde{\delta}_2^\perp(\bm{k,\hat{n}}) \\ + & \beta_{\mathcal{G}_2}(k,\mu)\tilde{\mathcal{G}}_2^\perp(\bm{k,\hat{n}}) + \beta_3(k,\mu)\tilde{\delta}_3^\perp(\bm{k,\hat{n}}),
\end{split}
\ee
where $\beta_i(k,\mu)$ are transfer functions which now depend on both $k$ and $\mu\equiv \bm{\hat k}\cdot \bm{\hat n}$. Here we define 
\be
\mathcal{G}_2^\parallel (\bm q) \equiv \hat{\bm n}_i \hat{\bm n}_j \frac{\partial_i\partial_j}{\nabla^2} \mathcal{G}_2 (\bm q) , 
\ee
which can be used to compute the shifted operator $\tilde{\mathcal{G}}_2^\parallel(\bm{k,\hat{n}})$ using Eq.~\eqref{eq:shifted_def}. Note that in redshift space the model error depends on both~$\k$ and~$\hat{\n}$ such that the full HI field can be written as
\be
\delta_\mathrm{HI}^s(\bm{k,\hat n}) = \delta_\mathrm{HI}^{s,{\rm model}}(\bm{k,\hat n}) + \epsilon_\mathrm{HI}^s(\bm{k,\hat n}) \;.
\ee

This model requires some explanations. While having a similar structure as in real space, Eq.~\eqref{eq:RSD_model} has new ingredients. In particular, the first line contains the Zel'dovich density in redshift space and shifted $\mathcal{G}_2^\parallel$ field. Note that both of these terms come without free coefficients. While these two terms can be in principle absorbed in other shifted operators (the same way as in real space $\delta_Z$ is absorbed in $\tilde\delta_1$, $\tilde{\mathcal G}_2$ and $\tilde{\mathcal G}_3$, see~\cite{Schmittfull}), we decide to keep them explicitly. Only this way it is guarantied that $\beta_1$, $\beta_2$, $\beta_{\mathcal G_2}$ and $\beta_3$ have the same limit when $k\to 0$ {\em independently} of $\mu$. This is important because it simplifies the analysis of the transfer functions in the low-$k$ limit and allows all $\mu$ bins to be combined. Such combination increases the signal-to-noise for the amplitude of each $\beta$ in redshift space, which in turn is important for comparison to real space results. This is another important difference in our work compared to previous literature, where the transfer functions are treated as constants independent of $\mu$ even without keeping $\tilde{\mathcal{G}}_2^\parallel$ in the model explicitly. 

One important consequence of explicitly keeping $\delta_Z^s$ and $\tilde{\mathcal{G}}_2^\parallel$ in the model is that the low-$k$ limits of the transfer functions in real and redshift space are not necessarily the same. One can show that when $k\to 0$, theoretical expectation is
\begin{align}
\label{eq:beta_r2zspace}
\beta_1^{\rm real} &= \beta_1^{\rm rsd} +1 ,  \\
\beta_2^{\rm real} &= \beta_2^{\rm rsd} ,  \\
\beta_{\mathcal{G}_2}^{\rm real} &= \beta_{\mathcal{G}_2}^{\rm rsd} + \frac 27 ,  \\
\beta_3^{\rm real} &= \beta_3^{\rm rsd} . 
\end{align}
We will test these relations once we fit the transfer functions in real and redshift space (see~\S\ref{sec:mocks} for comparison of this theoretical prediction and the best-fit values of transfer functions from simulations). 

In principle, one can compute the one-loop corrections for $\beta_1(k,\mu)$ in a similar way as in Eq.~\eqref{eq:beta1_model}, and we give the equations in Appendix~\ref{app:redshift_app}. However, with the precision available from a single box of Illustris simulation, testing the redshift-space distortions (RSD) model by fitting this transfer function would not be very informative. For this reason we will provide only a phenomenological fit to $\beta_1(k,\mu)$ in this paper, and leave a more detailed analysis for the time when larger simulation volumes will be available. All other transfer functions are expected to be constant on large scales, at leading order in perturbation theory.

\begin{figure*}[!ht]
\subfloat{\includegraphics[width=0.96\textwidth]{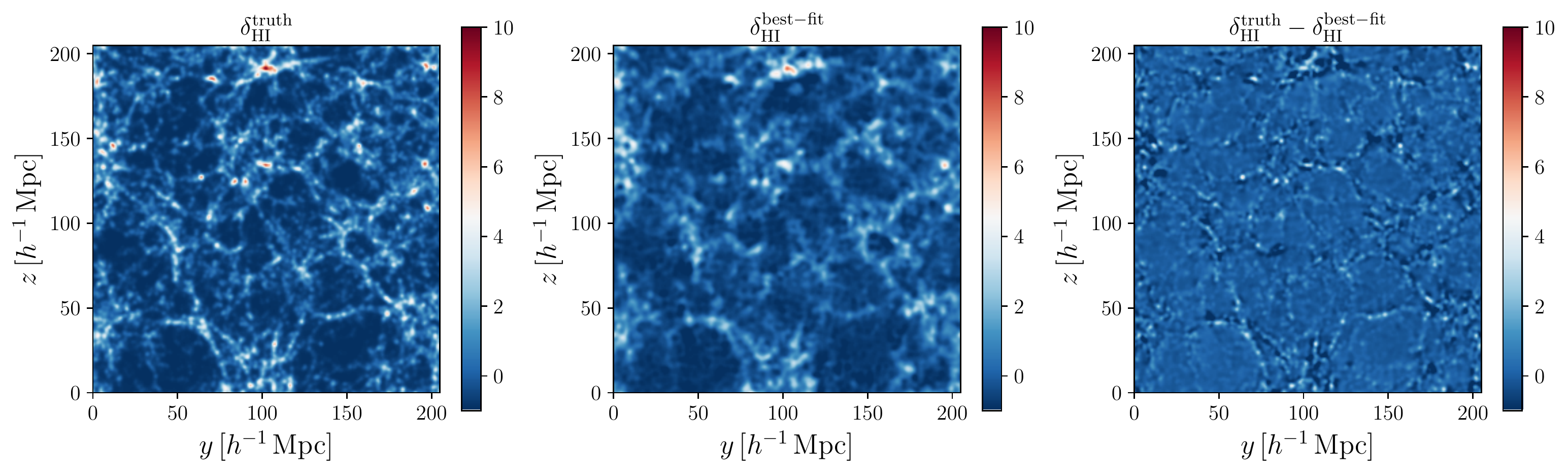}}\\
\subfloat{\includegraphics[width=0.96\textwidth]{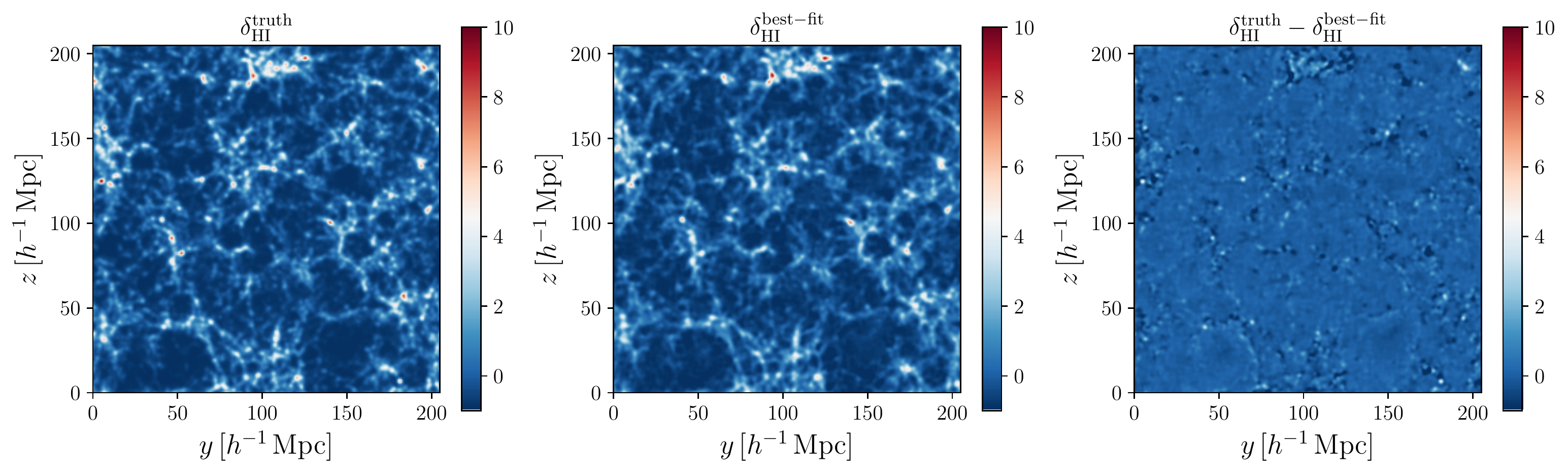}}\\
\caption{Real-space slices in the $y-z$ plane of the simulated HI overdensity field (left), best-fit cubic bias model (middle) and the residuals (right), at $z=0$ (\textit{top}) and $z=1$ (\textit{bottom}). All density fields are smoothed with a $R=1\Mpch$ 3D Gaussian filter, while the depth of each slice is $20\Mpch$.}
\label{fig:real_slices}
\end{figure*}

\subsection{Numerical implementation}
We first generate a Gaussian linear density field using the same initial conditions (IC) as were used to for the full hydrodynamical simulation (IllustrisTNG). We do it on a regular $256^3$ grid at the initial redshift $z_\mathrm{ini}=127$, using the linear power spectrum and random seed of TNG. We use the same $\Lambda$CDM cosmology as TNG: $\Omega_\mathrm{\Lambda}=0.6911$, $\Omega_\mathrm{M}=0.3089$, $\Omega_\mathrm{b}=0.0486$, $\sigma_8=0.8159$, $n_s=0.9667$ and $h=0.6774$, in agreement with the results from Planck \cite{Planck15}.

We generate a catalog of particles positioned on a regular $256^3$ grid within the simulation box ($L=205\hMpc$), which results in a particle separation $\Delta q = 0.8\hMpc$. At each particle's Lagrangian position $\bm{q}$ we compute the following: the displacement field $\bm{\psi}_1(\bm{q})$, $\delta_1(\bm{q})$, $\delta^2_1(\bm{q})$ and $\mathcal{G}_2(\bm{q})$. We then displace the particles by the displacement field $\bm{\psi}_1(\bm{q})$ and and place the displaced particles on the same regular grid using the cloud-in-cell (CIC) mass assigning scheme \cite{HockneyEastwood}. To obtain the shifted overdensity fields: $\tilde{\delta}_1$, $\tilde{\delta}_2$, $\tilde{\mathcal{G}}_2$ and $\tilde{\delta}_3$ we additionally weight displaced particles by the appropriate weights: $\delta_1(\bm{q})$, $\delta^2_1(\bm{q})$, $\mathcal{G}_2(\bm{q})$ and $\delta^3_1(\bm{q})$, respectively. We apply the same smoothing to $\delta_1$ when computing the shifted cubic term $\tilde{\delta}_3$ \cite{Schmittfull}.

In redshift space, we follow a similar procedure, except that the displaced particles are additionally shifted along the line of sight. We choose the simulation $z$-axis to be the line of sight ($\bm{\hat{z}})$ and for this choice the total displacement along the line of sight is then $(1+f)\bm{\psi}_{zz}$, where $f$ is the logarithmic growth rate \cite{Schmittfull_RSD}. Furthermore, we compute the ${\mathcal{G}}_2^\parallel$ term in Fourier space in the initial conditions as ${\mathcal{G}}_2^\parallel(\bm{k,\hat{n}})=\mu^2{\mathcal{G}}_2(\bm{k})$, before shifting it the usual way to obtain $\tilde{\mathcal{G}}_2^\parallel$.

\section{Simulated HI field}\label{sec:sims}
In this section we describe the simulations used to obtain the evolved HI field. We use the IllustrisTNG simulations \cite{TNGa,TNGb,TNGc,TNGd,TNGe}. We focus on the largest simulation box with highest resolution – TNG300-1, with a box size $L=205\hMpc$. We note that the main results in ref.\ \cite{PacoTNG} were obtained from a smaller simulation – TNG100-1 ($L=75\hMpc$) with higher mass resolution, which allowed to probe the HI distribution within smaller halos. However, for our work we need larger scales for two reasons. One reason is that the Zel'dovich displacement field receives a significant contribution from modes comparable/larger to the size of the smaller box, thus a larger box is needed to properly account for the large modes. Another reason is that we want to measure the performance of the model on large, BAO scales $\approx 100 \Mpch$ relevant for future 21cm IM surveys, which again requires a larger box. Furthermore, we shall focus on outputs at $z=0$ and $z=1$ as these are the key redshifts for several future surveys.

We compute the HI masses in post-processing following the approach from ref. \cite{PacoTNG}. We then assign the particles weighted by their HI masses on a regular $256^3$ grid using Cloud-in-Cell (CIC) mass assignment scheme \cite{HockneyEastwood}. We correct for the CIC window to obtain the final HI overdensity field in real space. Similarly, in order to obtain the HI overdensity field in redshift space, we apply the RSD shift along the line of sight prior to mesh assignment. We caution the Reader that the HI simulation results are not fully converged against resolution (see Appendix A in \cite{PacoTNG}).

In order to compare to the standard HI shot noise estimates, we make use of halo catalogs with total HI masses obtained from the same simulation TNG300-1. The halos were identified using Friends-of-Friends (FoF) finder and the total HI mass per halo is computed by summing the HI masses of all the particles within each FoF halo \cite{PacoTNG}.

\begin{figure*}[!ht]
\subfloat{\includegraphics[width=0.48\textwidth]{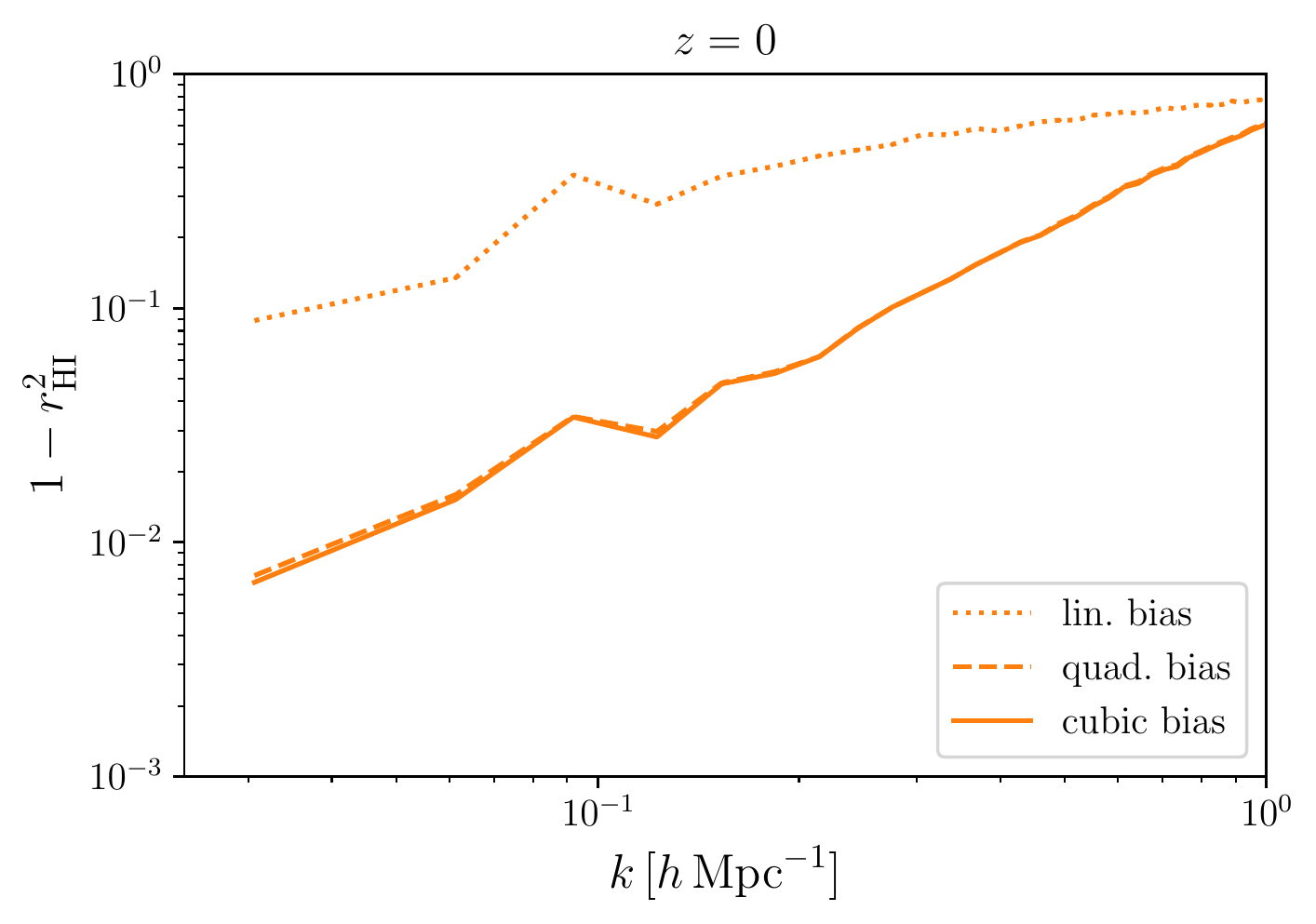}}
\subfloat{\includegraphics[width=0.48\textwidth]{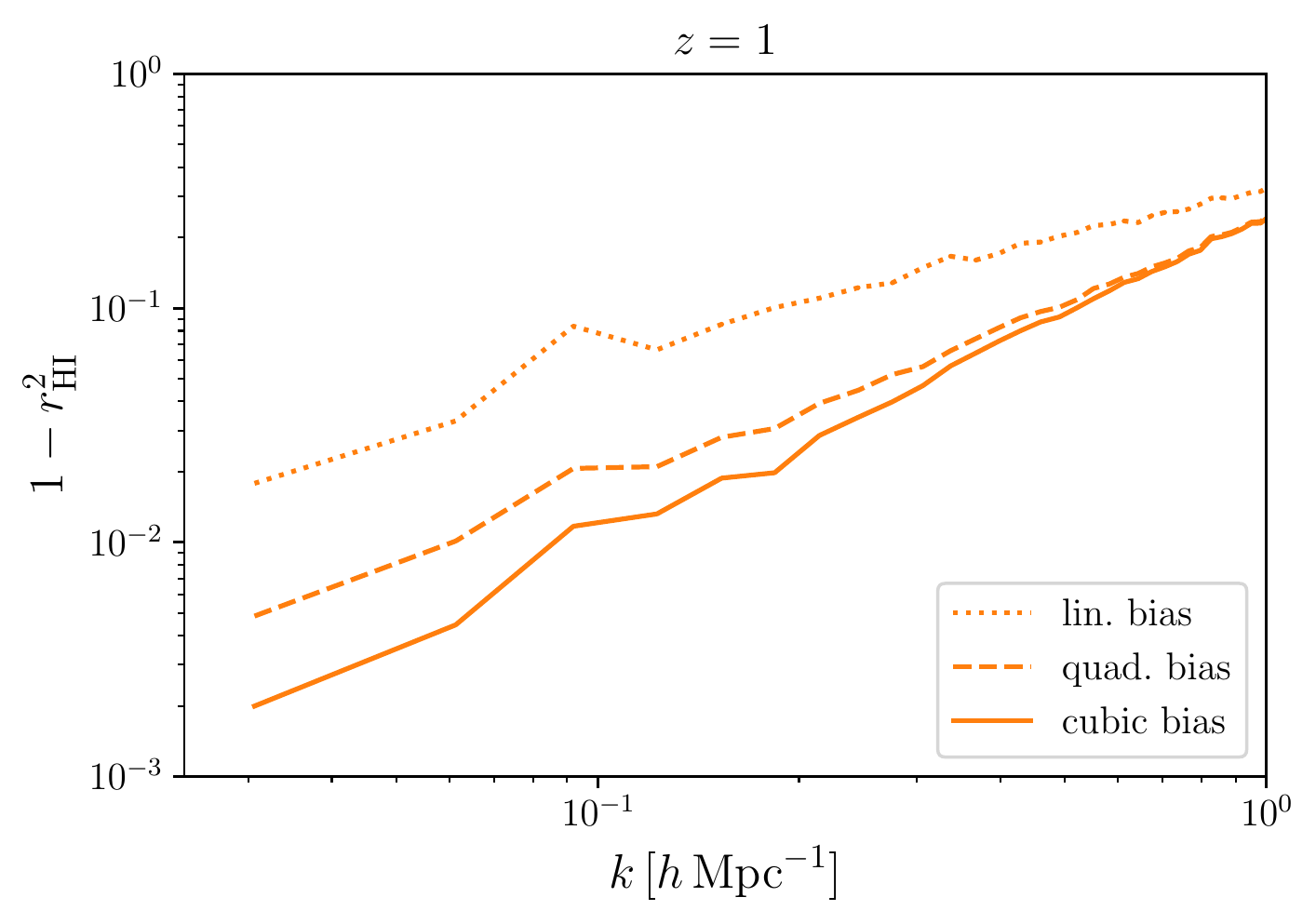}}
\caption{Relative error of the best-fit model $\Perr$ compared to the true HI power spectrum $P_\mathrm{truth}$. This coincides with $1-r_\mathrm{HI}^2$, where $r_\mathrm{HI}$ is the  cross-correlation coefficient between simulations and model. We show this quantity for different bias models: linear, quadratic and cubic, at $z=0$ (left) and $z=1$ (right). We find that using quadratic or cubic models significantly decreases the relative error compared to the linear bias model.}
\label{fig:rk_HI}
\end{figure*}

\begin{figure*}[!ht]
\subfloat{\includegraphics[width=0.48\textwidth]{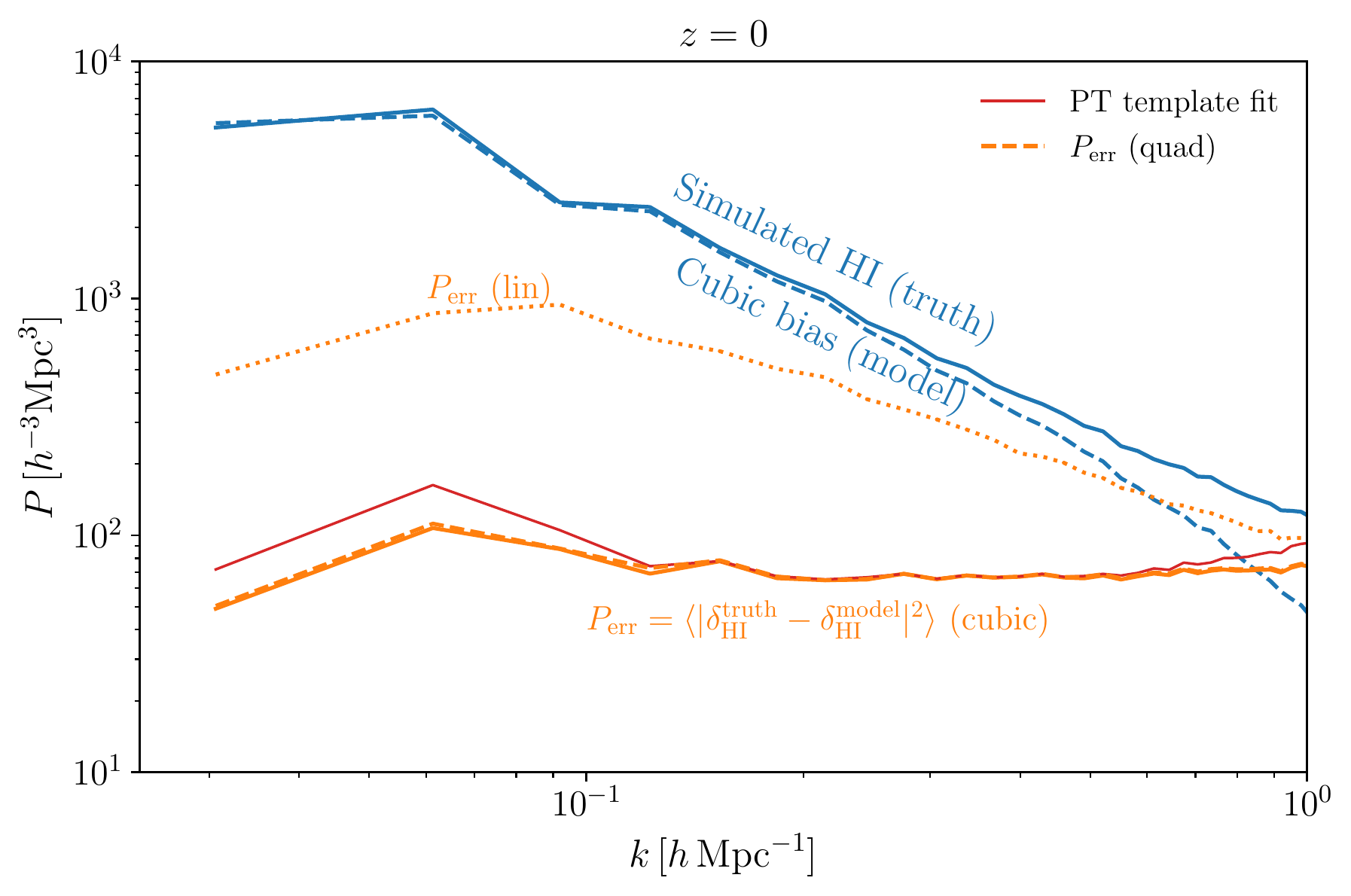}}
\subfloat{\includegraphics[width=0.48\textwidth]{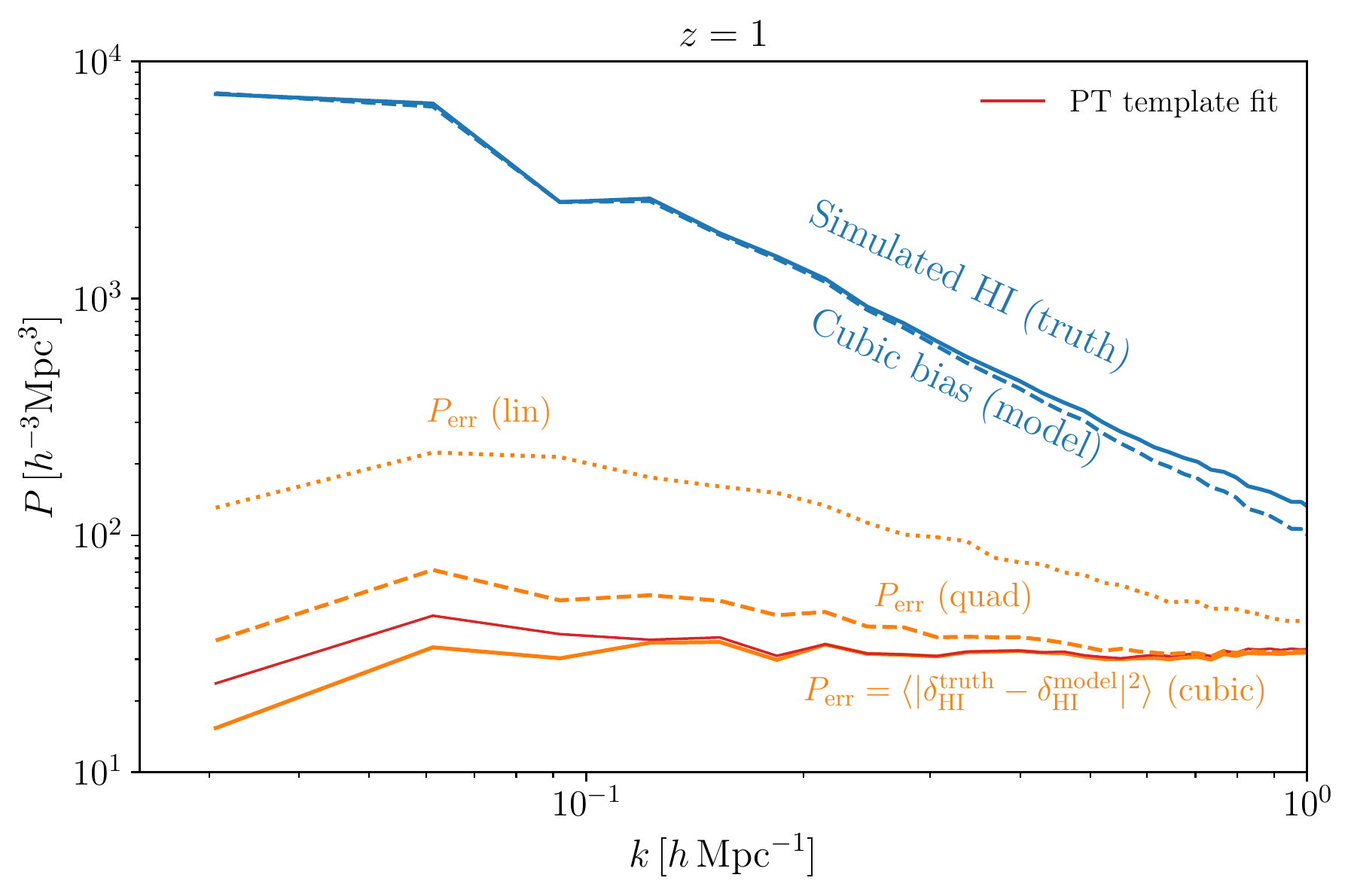}}
\caption{Real-space power spectrum: simulated (blue solid line), best-fit cubic model (blue dashed line) and error power spectra $\Perr$ at $z=0$ (\textit{left}) and $z=1$ (\textit{right}). We show $\Perr$ for different bias models used: cubic (solid orange line), quadratic (dashed orange) and linear (dotted orange line). We also show $\Perr$ when using the best-fit PT template based transfer functions for the cubic bias model (thin red line).}
\label{fig:real_pks}
\end{figure*}

\section{Results}\label{sec:results}

In this section we test how well the perturbative forward model matches the simulations and we present some results. Our main analysis is performed at the field level, where we fit the entire perturbative realization to simulations without restricting ourselves to any summary statistics.
The key object in this approach is the mean-square error/residual power spectrum, in general given by
\be
\Perr(\bm{k},\hat{\bm n}) \equiv \vev{|\delta_{\rm HI}^{\rm truth}(\bm{k},\hat{\bm n})-\delta_{\rm HI}^{\rm model}(\bm{k},\hat{\bm n})|^2}.
\ee
This definition implies that the error contains everything that is not included in the model and that therefore does not correlate with $\delta_{\rm HI}^{\rm model}$. Consequently,
\be
\label{eq:sum_of_model_and_noise}
P_{\rm HI}^{\rm truth} (\bm{k},\hat{\bm n}) = P_{\rm HI}^{\rm model} (\bm{k},\hat{\bm n}) + \Perr(\bm{k},\hat{\bm n}) \;.
\ee
In order to obtain the best fit model we minimize $\Perr$ in each $(k,\mu)$ bin. Since the model is linear in transfer functions, this minimization is equivalent to a linear regression in every bin. For the orthogonal basis of operators, the estimate of transfer functions is straightforward 
\be
\label{eq:transfer_best_fit}
\beta_i(k,\mu) = \langle\mathcal{O}^\perp_i(\bm{k},\hat{\bm n})\delta_{\rm HI}^{\rm truth}(\bm{k},\hat{\bm n})^*\rangle/\langle|\mathcal{O}^\perp_i(\bm{k},\hat{\bm n})|^2\rangle.
\ee
Note that in real space the $\Perr$ and the transfer functions $\beta_i$ are functions of the wavenumber amplitude $k$ only.

Before diving into the details, let us comment on general expectations for the behaviour of the transfer functions. On large scales, where the model is known to be correct, transfer functions are simply numbers and they do not depend on scale. On the other hand, deeply in the nonlinear regime where the model is completely wrong, transfer functions approach zero. This can be seen directly from Eq.~\eqref{eq:transfer_best_fit}, if one assumes that the truth does not correlate with the model on very small scales. In the intermediate regime, the transfer functions can be modeled by perturbation theory, and we will test these predictions in this section.

Let us also comment on the expected size of the error $\Perr$. Even if the model is perfect, we still expect the nonzero error for realistic biased tracers, coming from the sampling noise due to the finite number of objects in simulations or galaxy catalogs. Therefore, on large scales, we expect the amplitude of the noise to have a flat power spectrum, with the amplitude approximately equal to $1/\bar n$, where $\bar n$ is the mean number density of galaxies. On the other hand, approaching the nonlinear regime, the error power spectrum can become dominated by the true nonlinearities not captured by the model. These include higher order terms in perturbation theory as well as nonperturbative contributions such as 1-halo term. We will exploit this fact in redshift space to estimate the size of the nonlinear velocity dispersion for HI.

\subsection{Results in real space}\label{sec:real_space}

Let us begin presenting our results starting from real space. Minimizing the error power spectrum as explained above, we measure the transfer functions in all $k$ bins. In Fig.~\ref{fig:real_slices} we show a 2D slice of the simulated HI overdensity field in comparison to the best-fit cubic model and the residual at redshifts $z=0$ and $z=1$. We can see a general agreement between the simulation and best-fit model. Notable differences in the residuals are present at non-linear scales and match the positions of large overdensities. These differences are slightly more pronounced at later times, as expected from the nonlinear evolution. We note that a part of the difference is due to the accumulated projected error, given the thickness of the slice which is $20\;\hMpc$.

A more quantitative way to compare the two maps is to calculate the cross-correlation coefficient between the simulated HI field and the best-fit model:
\be
r_\mathrm{HI} \equiv \frac{\langle \delta_\mathrm{HI}^\mathrm{truth} \delta_\mathrm{HI }^\mathrm{best-fit*} \rangle}{(\langle|\delta_\mathrm{HI}^\mathrm{truth}|^2\rangle \langle|\delta_\mathrm{HI}^\mathrm{best-fit}|^2\rangle )^{1/2}}.
\ee
We show the quantity $1-r_\mathrm{HI}^2$ in Fig.~\ref{fig:rk_HI}, which can be shown to be equal to $\Perr/P_\mathrm{truth}$ using Eq.~\eqref{eq:sum_of_model_and_noise} (see Appendix B in ref.~\cite{Schmittfull}). It is clear that both for $z=0$ and $z=1$ the quadratic and cubic bias models have much better cross-correlation coefficient than the linear bias model, even on largest scales where the linear theory is supposed to work well. We will come back to this peculiar feature of HI clustering and discuss it in detail in Section~\ref{sec:HInoise}. It is important to note that the major source of disagreement between the simulations and the model on large scales is the sampling noise and not the failure of perturbation theory to predict the correct realization of the maps.

In order to see this more explicitly, we measure the best-fit power spectrum as well as the model error power spectrum. The results are shown in Fig.~\ref{fig:real_pks}. As previously mentioned, the model error power spectrum is expected to be flat if all the relevant higher-order terms are properly accounted for. Indeed, we find that using either the quadratic or the cubic model results in almost scale-independent~$\Perr$. On large scales we also find an excellent agreement between the power spectra in simulations and for the best fit cubic bias model. Note that the sum of the model and error power spectrum is equal to the simulated power spectrum by construction. 

\begin{figure*}[!ht]
\subfloat{\includegraphics[width=0.48\textwidth]{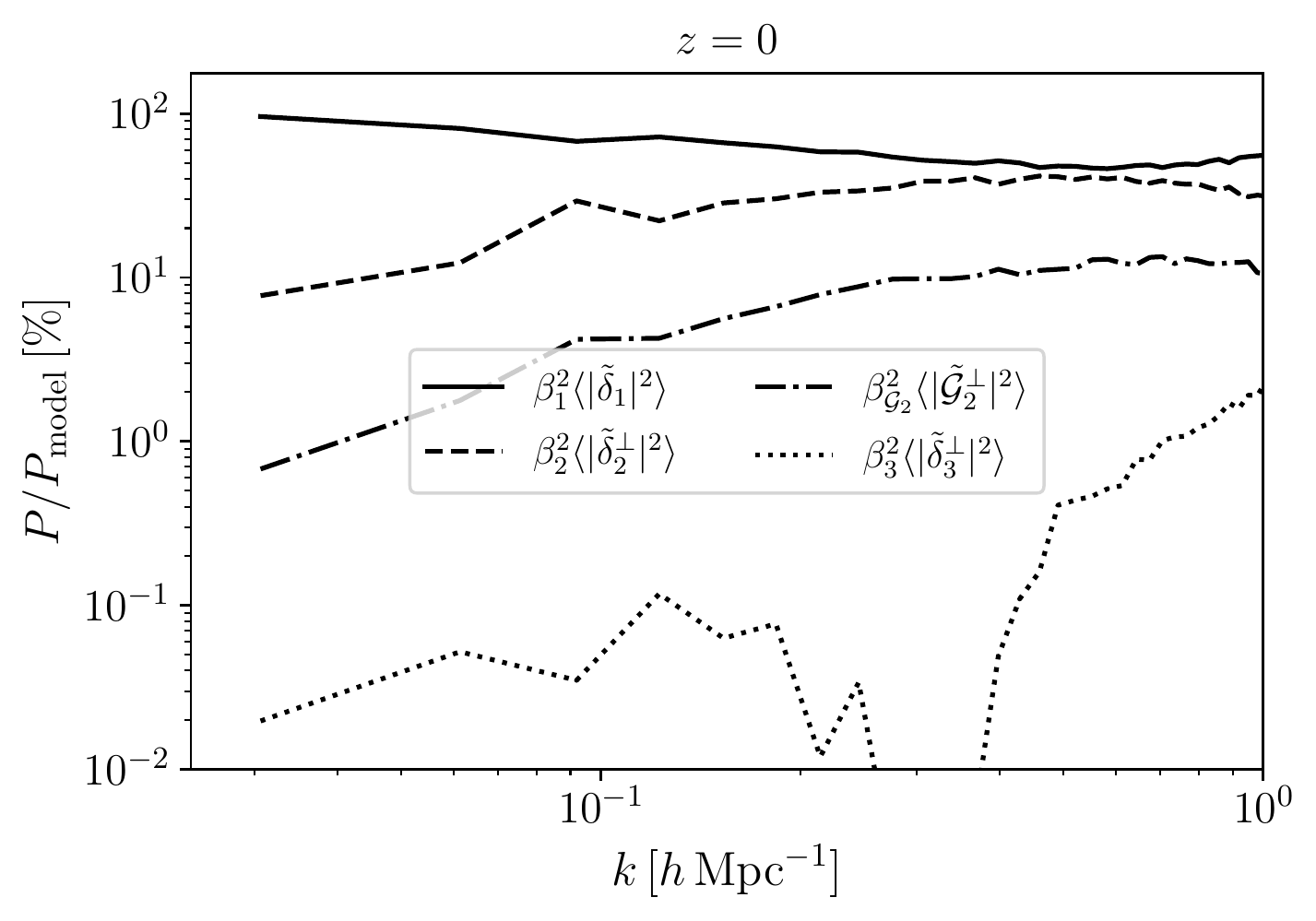}}
\subfloat{\includegraphics[width=0.48\textwidth]{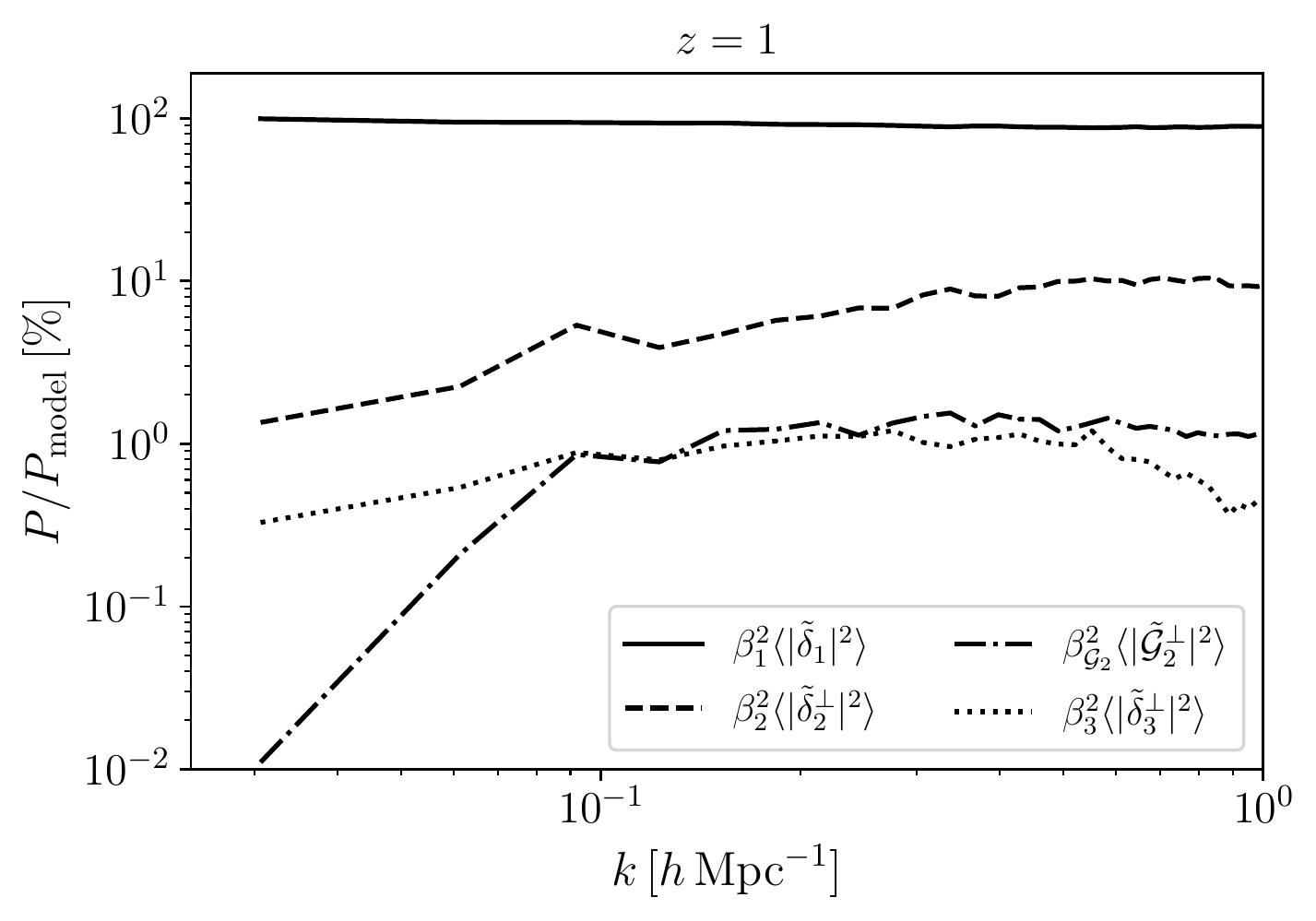}}
\caption{Relative contribution of different terms of the best-fit power spectrum using the cubic bias model in real space. Different panels correspond to different redshifts at $z=0$ (\textit{left}) and $z=1$ (\textit{right}). }
\label{fig:real_pk_rel_contr}
\end{figure*}

It is instructive to see how different nonlinear terms in the model contribute to the model power spectrum. In Fig.~\ref{fig:real_pk_rel_contr} we show the relative contributions to the best-fit power spectrum of each of the operators. We find that at both redshifts the linear term ($\tilde{\delta}_1$) dominates the signal. At $z=0$ the quadratic term ($\tilde{\delta}_2$) contributes $\sim10\%$ and more on small scales, while the tidal field ($\tilde{\mathcal{G}}_2$) is relevant at the level of $1-10\%$ on all scales. The picture changes at $z=1$, where the quadratic term contributes $1-10\%$ on all scales, while all higher order terms are below the sub-percent level. It is important to stress that the $\langle|\tilde{\delta}_2^\perp|^2\rangle$ contains a large contribution which has a flat power spectrum. Indeed, this is the reason why including the quadratic operators significantly reduces the error power spectrum compared to the simple linear bias model. Therefore, the fact that $\langle|\tilde{\delta}_2^\perp|^2\rangle$ is larger than other nonlinear terms and even comparable to $\langle|\tilde{\delta}_1|^2\rangle$ on small scales for $z=0$, should not be seen as a failure of perturbation theory, but rather as a success in describing the true dynamics of HI clustering at the field level.

\begin{figure*}[!ht]
\subfloat{\includegraphics[width=0.96\textwidth]{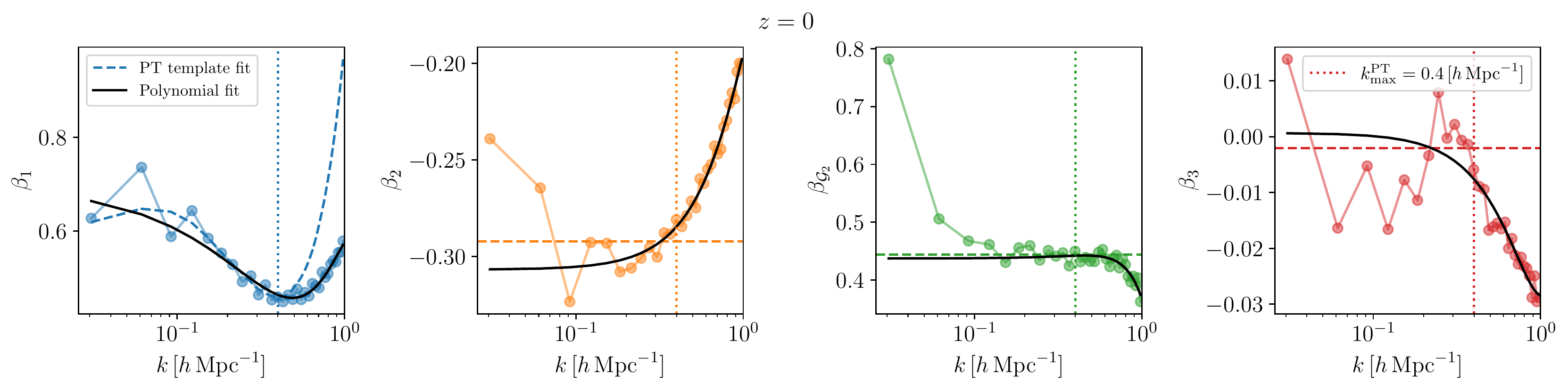}}\\
\subfloat{\includegraphics[width=0.96\textwidth]{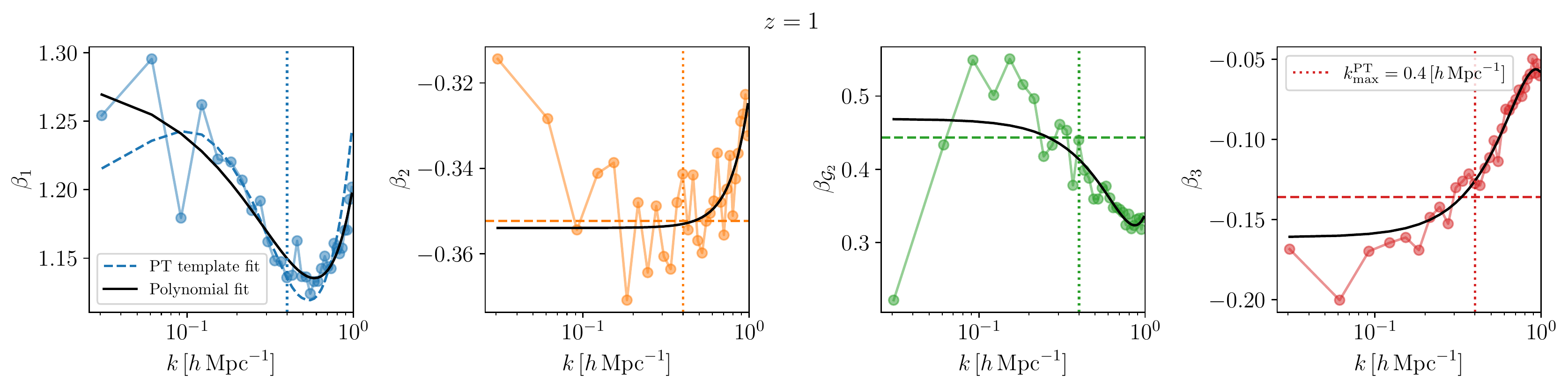}}
\caption{Transfer functions $\beta_i(k)$ in real space for the cubic bias model at at $z=0$ (\textit{top}) and $z=1$ (\textit{bottom}) (lines with different colors). We show the best-fit PT prediction (dashed lines) obtained using scales below $k_{\rm max}=0.4\hMpc$ (vertical dotted lines). We also show the fits obtained using the following polynomials : $\beta_1(k) = a_0 + a_1k + a_2k^2 + a_4k^4$, and $\beta_{i}(k) = a_0 + a_2k^2 + a_4k^4$ for higher order transfer functions (see table \ref{tab:polynomial_bestfits}) (black solid lines).}
\label{fig:best_fit_transfer}
\end{figure*}

\begin{figure*}[!ht]
\label{fig:rel_err_real}
\subfloat{\includegraphics[width=0.48\textwidth]{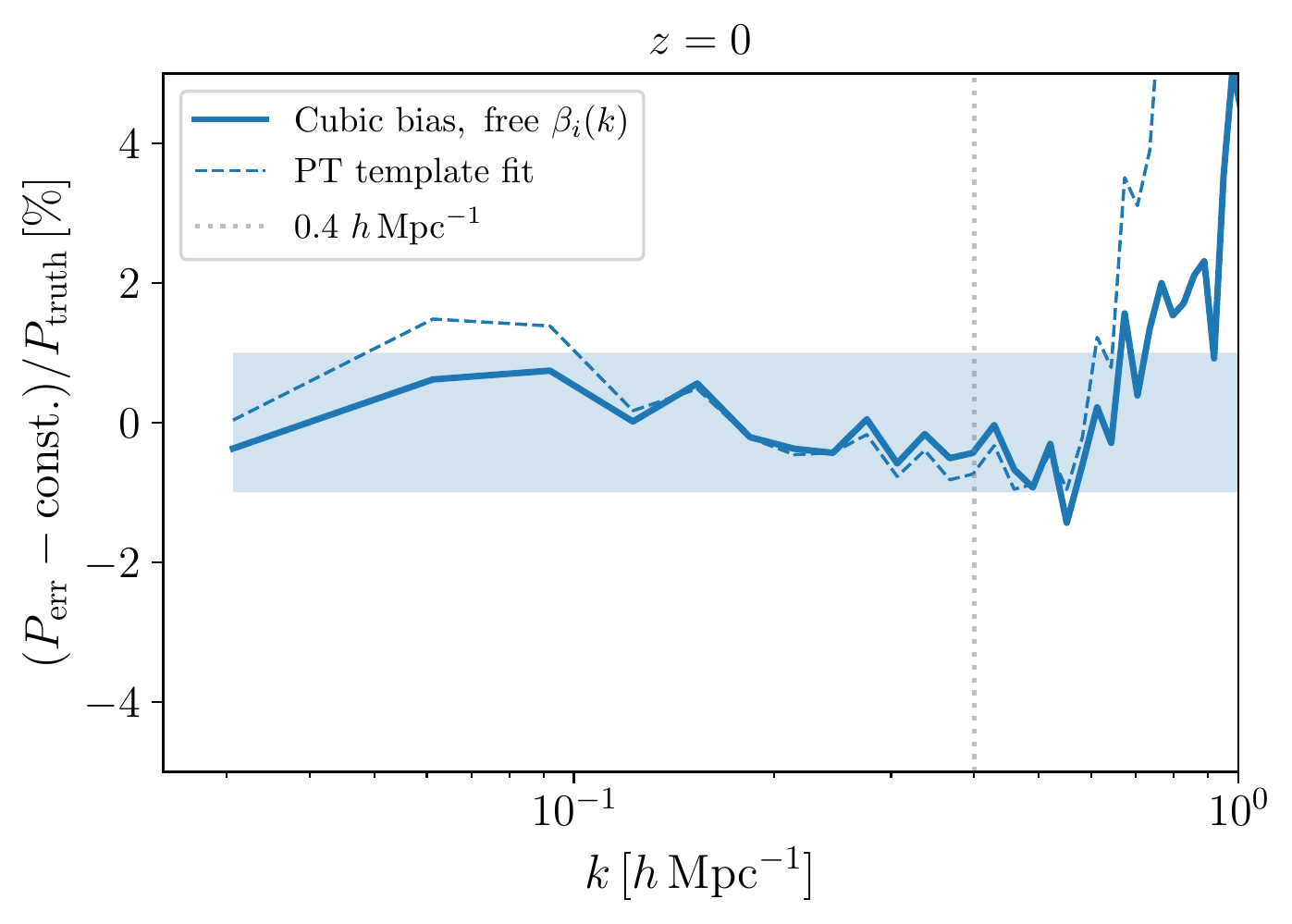}}
\subfloat{\includegraphics[width=0.48\textwidth]{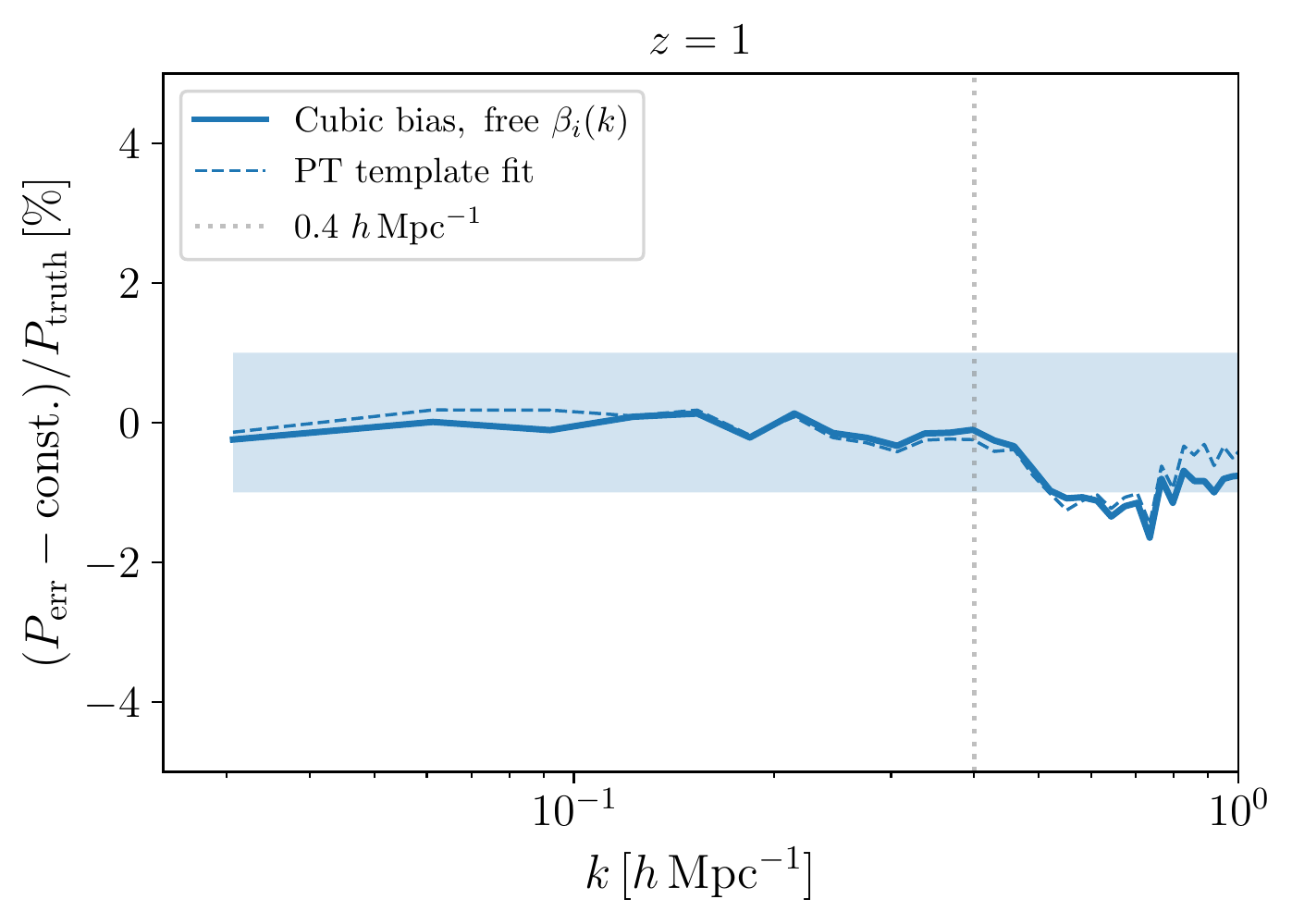}}
\caption{Fractional deviation of the error power spectrum $\Perr$ from constant using free transfer functions $\beta_i(k)$ (solid lines) and PT template fit with 6 free parameters (dashed lines), at $z=0$ (\textit{left}) and $z=1$ (\textit{right}). Note that this quantity is identical to the relative error of the perturbative model compared to the simulated HI power spectrum. The blue shaded region shows $\pm1\%$ of $P_\mathrm{HI}^\mathrm{truth}$.}
\label{fig:best_fit_transfer_Perr}
\end{figure*}

\begin{figure}[!ht]
\includegraphics[width=0.48\textwidth]{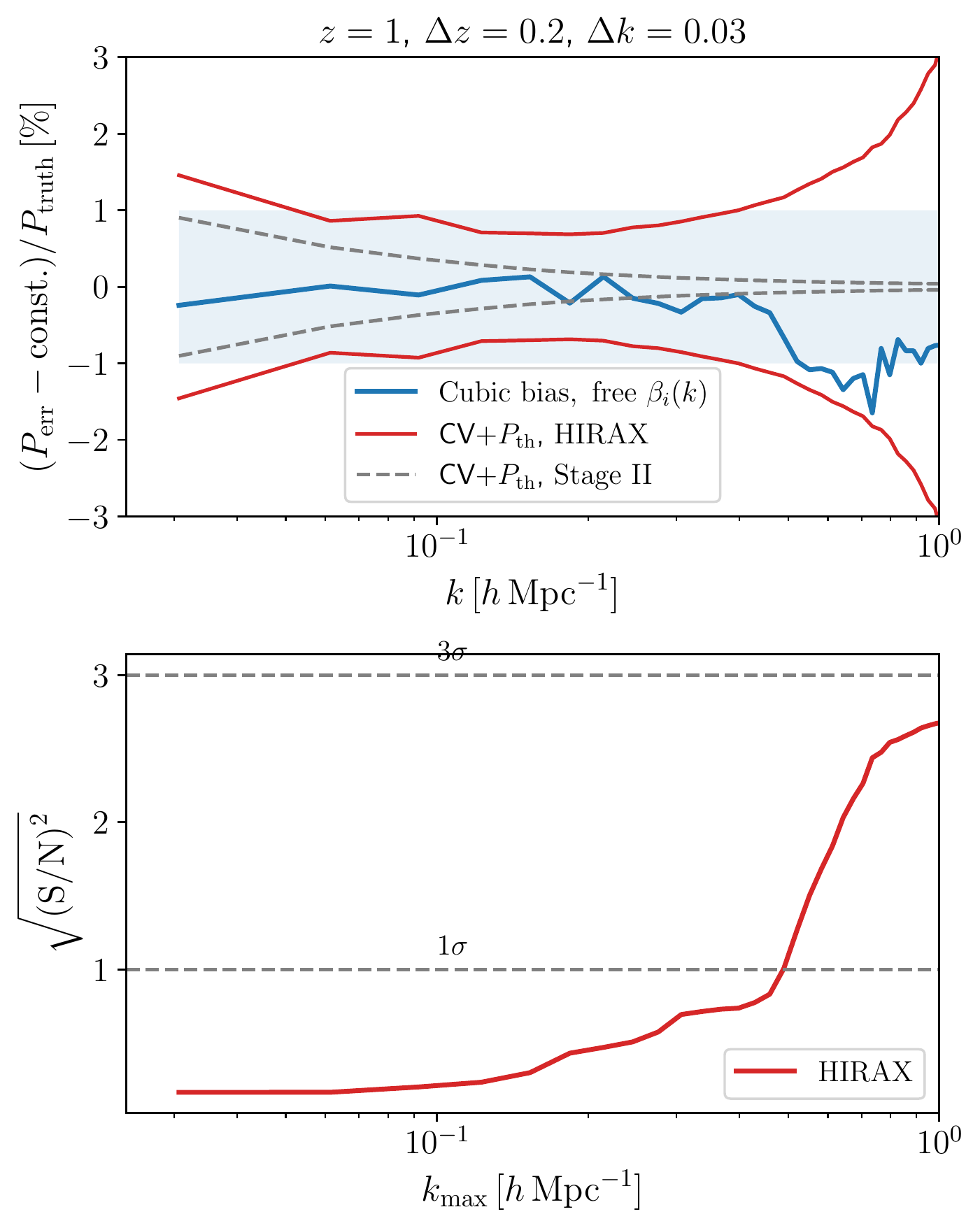}
\caption{\textit{Top panel}: Similar to Fig.~\ref{fig:best_fit_transfer_Perr} at $z=1$ but comparing to the expected performance of different 21cm IM surveys. The blue line is the relative error of the best-fit cubic bias model with free transfer functions. The solid red and dashed gray lines show the expected uncertainty after including the thermal noise and the cosmic variance for a HIRAX-like experiment and the Stage II experiment, respectively. The blue shaded region represents $\pm1\%$ of $P_\mathrm{HI}^\mathrm{truth}$. \textit{Bottom panel}: Cumulative signal-to-noise ratio squared $(\mathrm{S/N})^2$ for detecting deviation of the best-fit cubic bias model from the true HI power spectrum as a function of $k_\mathrm{max}$ for a HIRAX-like survey.}
\label{fig:Perr_const_surveys}
\end{figure}

\begin{figure*}[!ht]
\subfloat{\includegraphics[width=0.96\textwidth]{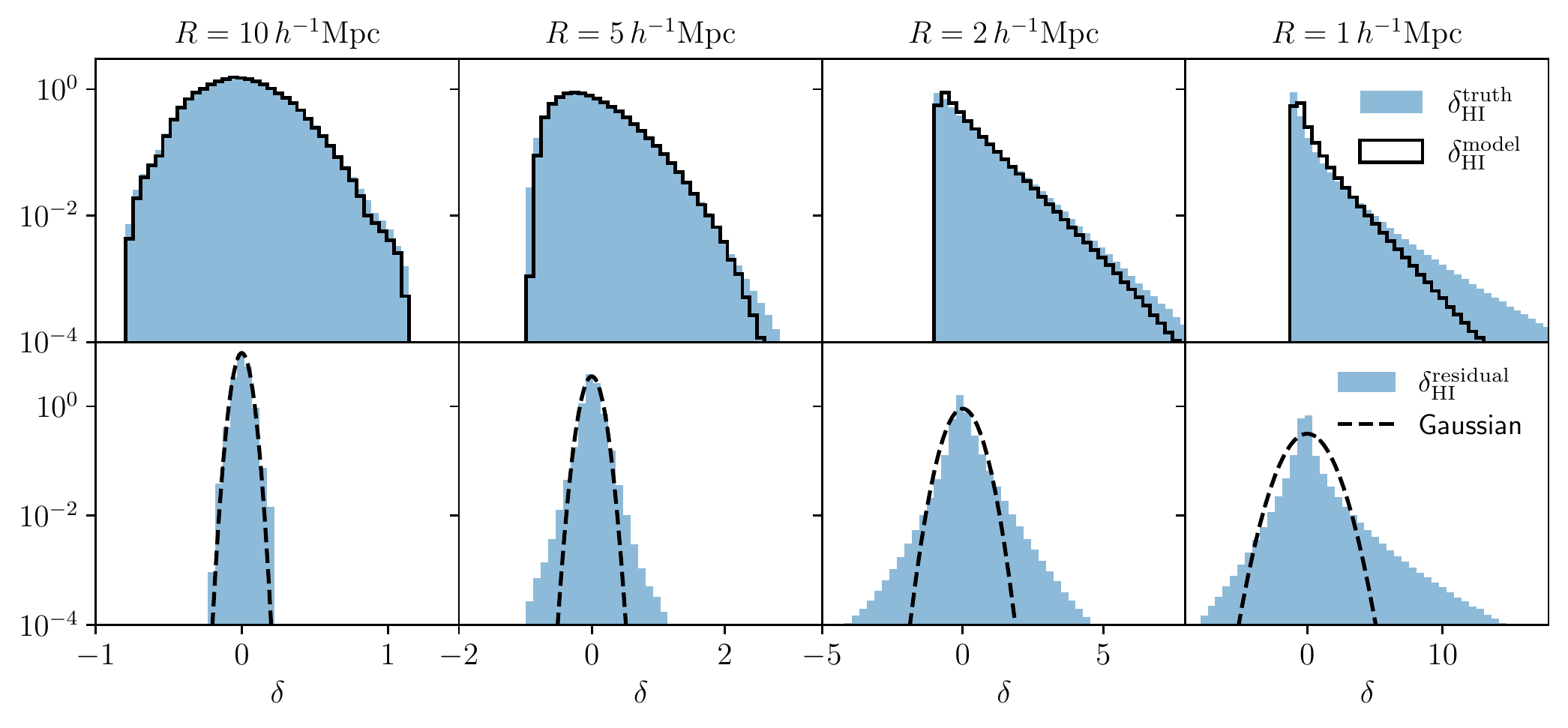}}\\
\subfloat{\includegraphics[width=0.96\textwidth]{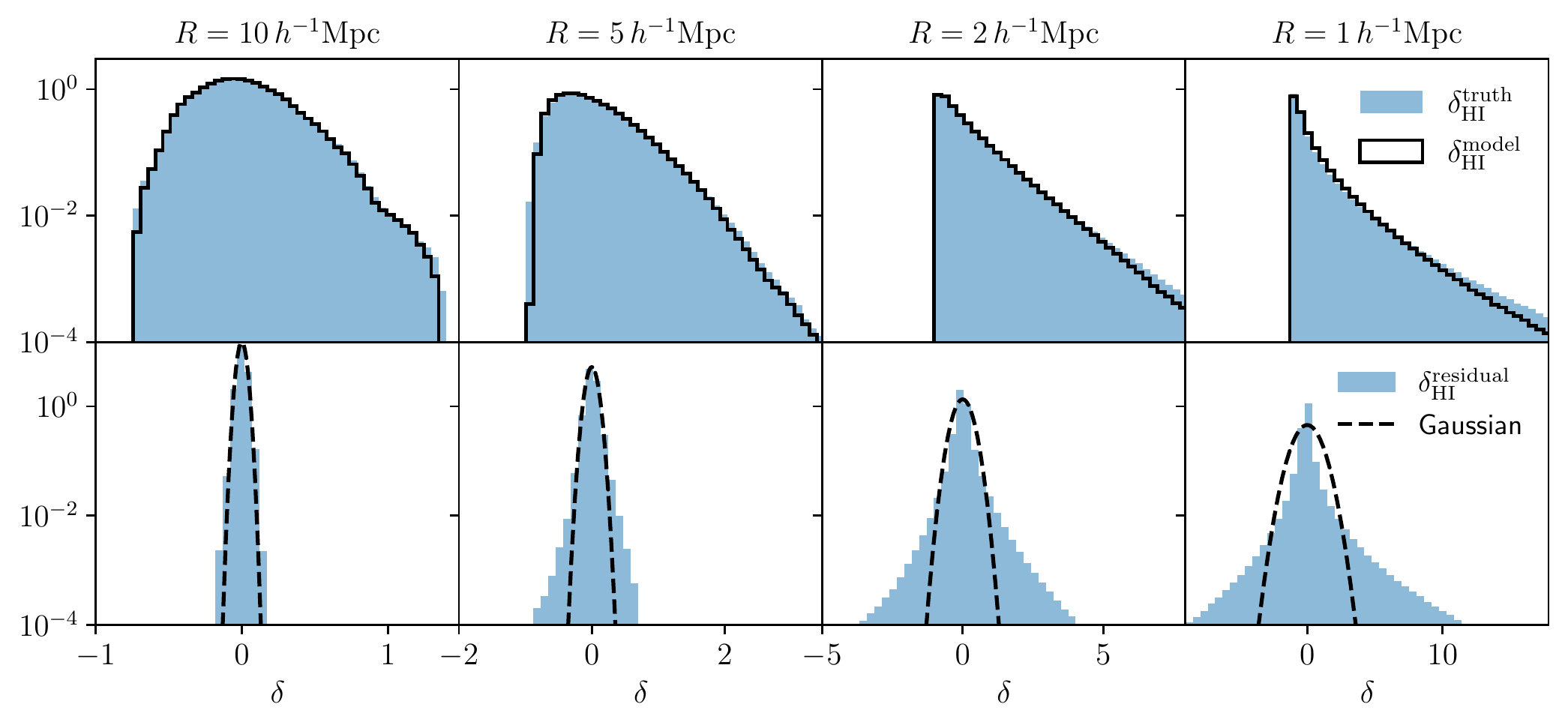}}
\caption{Histogram of the true, best-fit and the residual HI overdensity field using the cubic bias model at two redshifts: $z=0$ (\textit{top panel}) and $z=1$ (\textit{bottom panel}). Different panel columns correspond to applying a 3D Gaussian filter to residuals with varying smoothing scale $R$. Top rows in each panel show the simulated (blue) and the best-fit model (black lines) HI overdensity distribution. We find that the (cubic) bias model provides a good description of the simulated HI one-point pdfs, especially on larger smoothing scales. Bottom rows show the best-fit field level residuals (blue) compared to a Gaussian distribution with matching variance and zero-mean. Using larger smoothing scales $R$, the residuals (or model errors) are nearly Gaussian, with larger deviations present in a small fraction of pixels.}
\label{fig:hist}
\end{figure*}

\subsection{Best-fit transfer functions}\label{sec:best_fit_transfer}
So far we have compared the best-fit perturbative forward model and simulations and found an excellent agreement between the two. However, the transfer functions that we used to minimize the mean-squared error have been completely free in each $k$-bin. One may wonder if such large freedom is justified. In other words, if the transfer functions are very scale-dependent or cannot be predicted in perturbation theory, this would undermine the success of the perturbative forward modeling. 

In this section we address this issue by comparing the measured best-fit transfer functions to perturbation theory predictions. As we have already discussed, below the nonlinear scale the transfer functions should be described by a handful of free (bias) parameters. We use the model from~\S \ref{sec:real_space_model} to test this prediction and measure the bias parameters. In this model only the first transfer function $\beta_1$ is scale-dependent, while others are treated as constants. As this model is valid only in the perturbative regime, for the purposes of comparison we include only scales below $k<0.4\hMpc$. We fit jointly all transfer functions $\beta_i(k)$ weighting each $k$ bin by $k$ to account for the different number of modes. The results are shown in Fig.~\ref{fig:best_fit_transfer}, where dashed lines represent the best-fit curves, while the best-fit values of parameters are provided in Table~\ref{tab:transfer_function_best_fit_values}. We also perform a convergence test with respect to our grid size choice and show that our results are converged in Appendix~\ref{app:grid_convergence}.

\begin{table}[!ht]
    \centering
    \begin{tabular}{c|c|c}
    $z$ & 0 & 1\\
    \hline
    $b_1$  & 0.600 & 1.204\\
    $c_s^2$ & 0.316 & 0.338\\
    $b_2$  & -0.292 & -0.352\\
    $b_{\mathcal{G}_2}$  & 0.444 & 0.443\\
    $b_{\Gamma_3}$  & -0.826 & -1.361\\
    $b_3$ & -0.002 & -0.136\\
    \end{tabular}
    \caption{Best-fit transfer function parameters from the PT model at different redshifts obtained using $k_\mathrm{max}=0.4\hMpc$.}
    \label{tab:transfer_function_best_fit_values}
\end{table}

As we can see, analytical prediction provides a decent fit for the transfer functions on scales~$k<0.4\hMpc$. Note that the points in the first few bins have a large scatter, as a consequence of the numerical noise given that we use only a single realization in simulations. This scatter could in principle be reduced by averaging over several realizations of the simulated HI field or using simulations with much larger volumes. It is also instructive to look at the values of best-fit parameters. For the consistency of the theory, the bias parameters measured in transfer functions should have the same values as those inferred using other methods, such as separate universe simulations~\cite{Lazeyras...02..018L,2018JCAP...07..029A,2020JCAP...12..013B}. While the direct comparison with the biases for dark matter halos is not fully appropriate for HI, we can see from Table~\ref{tab:transfer_function_best_fit_values} that the values of $b_1$, $b_2$ and $b_3$ are close to the expectation for halos with~$M_h\sim 10^{11}\ [h^{-1} M_\odot]$, which dominantly contribute to the HI signal. This is yet another confirmation that our perturbative framework is consistent. 

Having obtained the best-fit PT template model for transfer functions, we are able to construct an approximate HI field using shifted operators and best-fit $\beta_i(k)$ model. We test the performance of such an approximate model, hereafter PT template model, by measuring $P_\mathrm{err}$. The results are shown in Fig.~\ref{fig:real_pks} in thin red line. We find that using only 6 free parameters gives results similar to the case of free transfer functions in each $k$-bin, albeit with somewhat larger error on small scales. Note that differences in the error power spectrum on large scales are a consequence of the numerical noise and the fact that we fit the noisy data with a smooth function, unlike for the free transfer functions case where all bins are independent. 

Finally, in Fig.~\ref{fig:best_fit_transfer_Perr} we show the fractional deviation of $\Perr$ from a constant when using PT template model. In comparison with the free transfer functions, the results show similar behaviour albeit with a larger scatter. As we already explained, this scatter is a result of the numerical noise. The fractional deviation of $\Perr$ from a constant can be also seen as the relative error on the overall power spectrum model
\be
\frac{\Perr - \mathrm{const.}}{P^\mathrm{truth}_\mathrm{HI}} = \frac{P^\mathrm{truth}_\mathrm{HI}-( P^\mathrm{model}_\mathrm{HI} + \mathrm{const.})}{P^\mathrm{truth}_\mathrm{HI}} .
\ee
The value of the constant is chosen as an average plateau of $\Perr(k)$ on $k=0.1-0.3\hMpc$.
We can see that the perturbation theory model describes the power spectrum well for $k<0.4\; \hMpc$, with sub-percent relative error. Due to the noise in the measurements coming from a single realization it is hard to estimate the true amplitude of the error as a function of scale. We leave this for future work when more simulation boxes or larger volumes will be available. Let us only note that on general grounds we expect that the error power spectrum to be scale-dependent around the nonlinear scale, with relative corrections of the order of $(k/k_{\rm NL})^2$. Including such terms in the error model would further improve the agreement with the simulated HI power spectrum.

\begin{figure*}[!ht]
\subfloat{\includegraphics[width=0.96\textwidth]{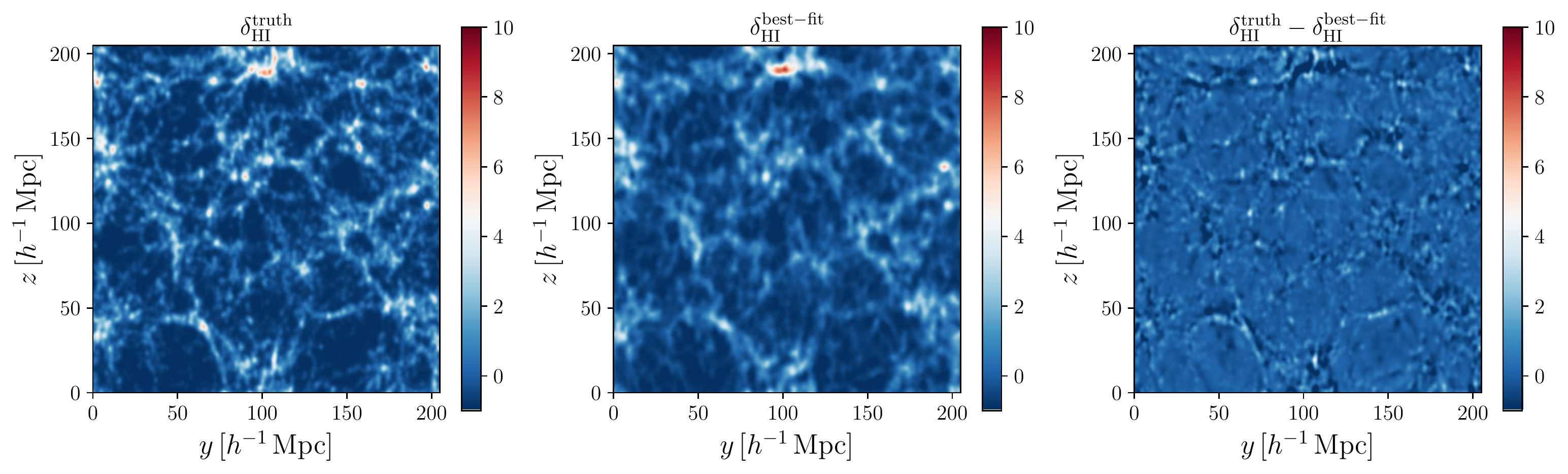}}\\
\subfloat{\includegraphics[width=0.96\textwidth]{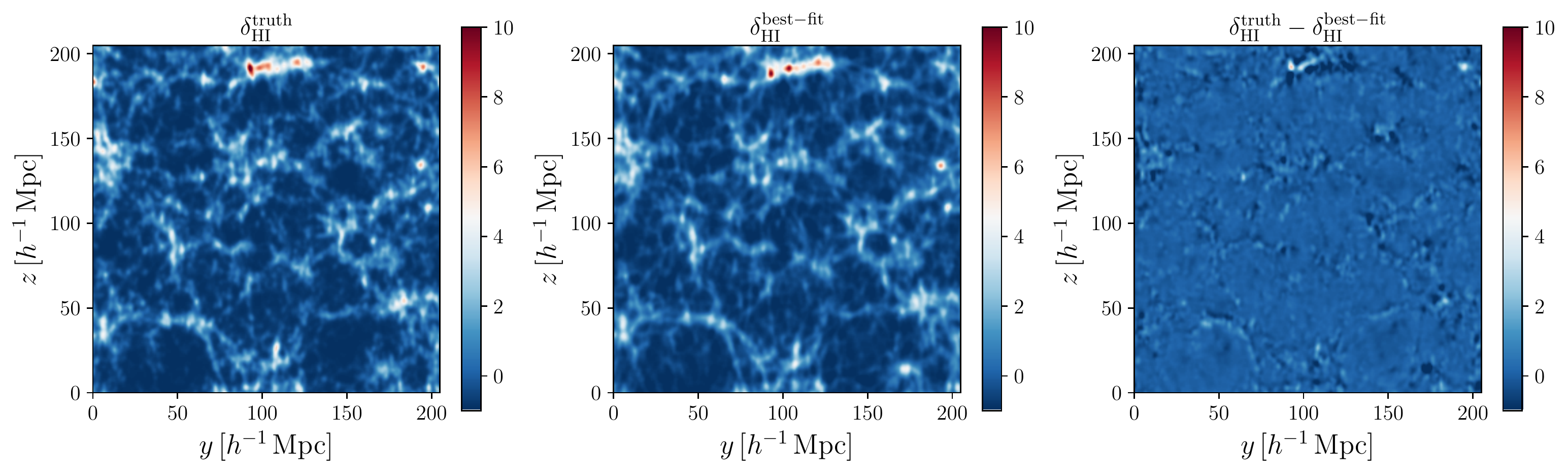}}
\caption{Similar to Fig.~\ref{fig:real_slices}, but now in redshift-space: Real-space slices in $y-z$ plane of the simulated HI overdensity field (left), best-fit cubic bias model (middle) and the residuals (right), at $z=0$ (\textit{top}) and $z=1$ (\textit{bottom}). All density fields are smoothed with a $R=1\Mpch$ 3D Gaussian filter, while the depth of each slice is $20\Mpch$. The line of sight is along $\hat z$ direction.}
\label{fig:rsd_slices}
\end{figure*}

\subsection{21cm IM surveys}\label{sec:im_surveys}
It is useful to contrast the performance of the perturbative model that we tested in the previous section to the expected noise levels from future 21cm IM surveys. These surveys are typically split into two main categories: single dish and interferometric experiments. The main difference between these is the range of scales they probe, with the former mainly probing large scales due to the size of the beam especially at higher redshifts, while the latter are more sensitive to smaller scales. As we are interested in testing the performance in the non-linear regime, we focus on the interferometric approach. We follow the experimental setup similar to those of HIRAX \cite{HIRAX} and Stage-II survey such as PUMA \cite{PUMA}. These surveys will consist of square, compact arrays of respectively $32\times32$ and $256\times256$ dishes with the diameter of $D_\mathrm{dish}=6\rm{m}$.

In this paper we consider an idealized setup, in which we neglect the impact of foreground cleaning \cite{Shaw14,Shaw15,Byrne, ska_foreground}, the foreground wedge \cite{wedge_parsons, wedge_pober,wedge_liu_1,wedge_liu_2} and assume an idealized measurement. Even though this is not realistic, our main purpose is to demonstrate that even with such optimistic assumptions, the standard one-loop perturbative model is sufficiently accurate for upcoming 21cm IM surveys. Having this in mind, the total error in the power spectrum measurements depends only on the cosmic variance and thermal noise. We first focus on instrumental thermal noise. Temperature fluctuations from the instrument and its surrounding, as well as the sky temperature fluctuations, add power into the visibility measurements. The power spectrum of the observed 21cm signal contains this thermal noise power spectrum \be
P_{21}(k,\mu) = P_\mathrm{HI}(k,\mu) + P_\mathrm{thermal}(k,\mu). 
\ee
We estimate it following the approach in refs.\ \cite{Bull,highz21cm,Synergies}
\be
\label{eq:Pthermal}
P_\mathrm{thermal} = \frac{T^2_\mathrm{sys}(z)X^2(z)Y(z) \lambda_{21}^4(z) S_{21}}{A^2_\mathrm{eff}\mathrm{FOV}(z)t_0 n_\mathrm{pol}n(\mathbf{u},z)}.
\ee
In the expression above the system temperature $T_\mathrm{sys}$ is the sum of  amplifier noise, background sky and ground temperatures, i.e.\ $T_\mathrm{sys} = T_\mathrm{amp} + T_\mathrm{sky} + T_\mathrm{gnd}$. We assume the following \cite{PUMA}: $T_\mathrm{amp}=62\ \mathrm{K}$, $\,T_\mathrm{gnd}=33\ \mathrm{K}$ and
\be
T_\mathrm{sky}=2.7 + 25\left(\frac{\nu_{21}(z)}{400\ \mathrm{MHz}}\right)^{-2.75}\ \mathrm{K},
\ee
where $\nu_{21}(z)=\nu_0/(1+z)$ is the observed frequency of the HI line emitted at redshift $z$, with $\nu_0=1420\ \mathrm{MHz}$ being the rest-frame frequency of the 21cm line. Similarly, $\lambda_{21}(z)=\lambda_0(1+z)$ is the observed wavelength of the HI emission from redshift $z$, with $\lambda_0=21\ \mathrm{cm}$. $X$ is the comoving radial distance and $Y(z)=\frac{c(1+z)^2}{\nu_0 H(z)}$. These terms are used to convert the angular and frequency space to the physical space. The total survey area $S_{21}$ is assumed to be $15000$ and $20000\ \mathrm{deg}^2$ for HIRAX and Stage-II surveys, respectively. Furthermore, we assume an effective area per beam given by $A_\mathrm{eff}=\pi(D_\mathrm{dish}/2)^2$ and each dish having an isotropic primary beam with the field of view $\mathrm{FOV}=(1.22\lambda_{21}(z)/D_\mathrm{dish})^2$. For the number density of baselines $n(\mathbf{u},z)$ we an approximation from ref.~\cite{PUMA} using the parameters of a square array, where $\mathbf{u}=\mathbf{L}/\lambda_{21}$ is the baseline separation vector between two antennae in units of the observed HI wavelength. Finally, we assume $n_\mathrm{pol}=2$ as the number of polarisation per antennae and the observing time of $t_0=1\ \mathrm{yr}$.

The power spectrum of the thermal noise from Eq.~\eqref{eq:Pthermal} is scale-dependent through an uneven distribution of baselines which measure fluctuations on different spatial scales. This makes $P_\mathrm{thermal}$ dependent on both $k$ and $\mu$. In particular, the lack of baselines with large separations results in high thermal noise power on small scales. In this way the thermal noise we include accounts for the availability of scales each survey will be able to probe.

The second component of the error in our idealized setup is the cosmic variance. For concreteness we imagine a slice of a survey centered at $z=1$ with the survey area $S_{21}$. The Gaussian contribution to the covariance matrix is given by
\be
C_{k,k'} = \frac{(2\pi)^3}{V_\mathrm{survey}}\frac{P_{21}^2}{2\pi k^2 \Delta k},
\ee
where $V_\mathrm{survey}$ is the comoving volume of the slice we consider and $P_\mathrm{HI}$ is the HI power spectrum. We choose a redshift bin width of $\Delta z=0.2$ and $\Delta k\approx 0.03 \hMpc$, the same $k$ binning that we have used throughout the paper.

The expected uncertainty from the combined cosmic variance and thermal noise for the two different setups that we consider is shown in the top panel of Fig.~\ref{fig:Perr_const_surveys}, with the same $k$-binning as we used for $P_\mathrm{HI}$. To obtain only the $k$-dependent part of the thermal noise power spectrum we compute the monopole part of $P_\mathrm{thermal}$ in each $k$-bin. We can see that the theoretical uncertainties are within the observational noise in the perturbative regime. For a HIRAX-like survey this is true even on smaller scales. A more quantitative way to test the significance of the deviation of theory from simulations is to calculate the signal to noise for this difference give the error bars. We plot such cumulative signal to noise for a HIRAX-like survey at the bottom panel of Fig.~\ref{fig:Perr_const_surveys}. Perturbation theory model is clearly consistent with the data for $k<0.4\;\hMpc$, and on smaller scales the significance of the difference compared to the truth never exceeds $3\sigma$ level due to increasing thermal noise. In a realistic data analysis, where cosmological parameters are varied as well, we expect this difference to be even smaller. Therefore, this result can be also interpreted as a hint that the current level of theoretical modeling discussed in this paper is sufficient to analyse the data from the upcoming HI surveys such as HIRAX.

We would like to emphasise that these results should be taken with the grain of salt. Given a relatively small volume of the IllustrisTNG box and a single realization, the numerical noise in the transfer functions and the best-fit model, clearly visible in Fig.~\ref{fig:Perr_const_surveys}, can impact the signal-to-noise estimate. For this reason we do not explicitly show the cumulative signal-to-noise for the Stage II surveys that have much smaller error bars. Also, note that our choice of $\Delta z$ in this exercise is somewhat arbitrary. In the real data analysis the choice of $\Delta z$ will be driven by the compromise between increasing the cosmological volume in each $z$-bin and probing the redshift evolution across the bin. While specific choices in the future are likely to be different, we follow what is done in the galaxy power spectrum analysis of the most recent galaxy surveys such as BOSS~\cite{BOSS}. 

\subsection{One-point probability distributions}\label{sec:1pt_pdfs}

So far we have been mainly focusing on the power spectrum as the standard observable used in cosmological data analyses. However, as we pointed out in the introduction, one of the advantages of the field level method is that it allows to predict other, very different summary statistics. As an illustration we will focus on the one-point probability distribution function (PDF). One-point PDF of HI from TNG100-1 has also been studied in ref. \cite{Countsincells_TNG}. Here we test our perturbative prediction for the one-point PDF by comparing the histograms of smoothed density fields from simulations and the best-fit cubic bias for various smoothing scales. The results are shown in Fig.~\ref{fig:hist}.

For large smoothing scales the agreement between the two is rather good, particularly at $z=1$, even for relatively large overdensities. Focusing on smaller smoothing scales, we find that the best-fit model tends to over-predict the number of pixels near the mean density ($\delta\sim 0$) and under-predicts the number of highly overdense pixels. This is very much in line with what is expected from perturbation theory, as it is unable to account for compactness of highly non-linear objects, such as halos. In conclusion, this is yet another evidence that the perturbative description of the HI field works as expected, predicting the 1-point PDF correctly, including the proper shape of the distribution even in the tails, for smoothing scales as small as $R=5\;\Mpch$. 

\subsection{Redshift-space}\label{sec:redshift_space}
Let us finally turn to redshift space. We can in principle repeat every analysis we did for real space and they would lead to similar results. In this section we briefly present the most important tests of the perturbative forward model---comparison of theory and simulations at the level of the map and power spectrum.

In Fig.~\ref{fig:rsd_slices} we show a 2D slice of the simulated HI overdensity field in redshift space in comparison to the best-fit cubic model and the residuals at redshifts $z=0$ and $z=1$. As in the case of real space, we can see a general agreement between the simulation and the best-fit model. Notable differences in the residuals are again more pronounced on small scales and they match the locations of large overdensities.

\begin{figure*}[!ht]
\subfloat{\includegraphics[width=0.48\textwidth]{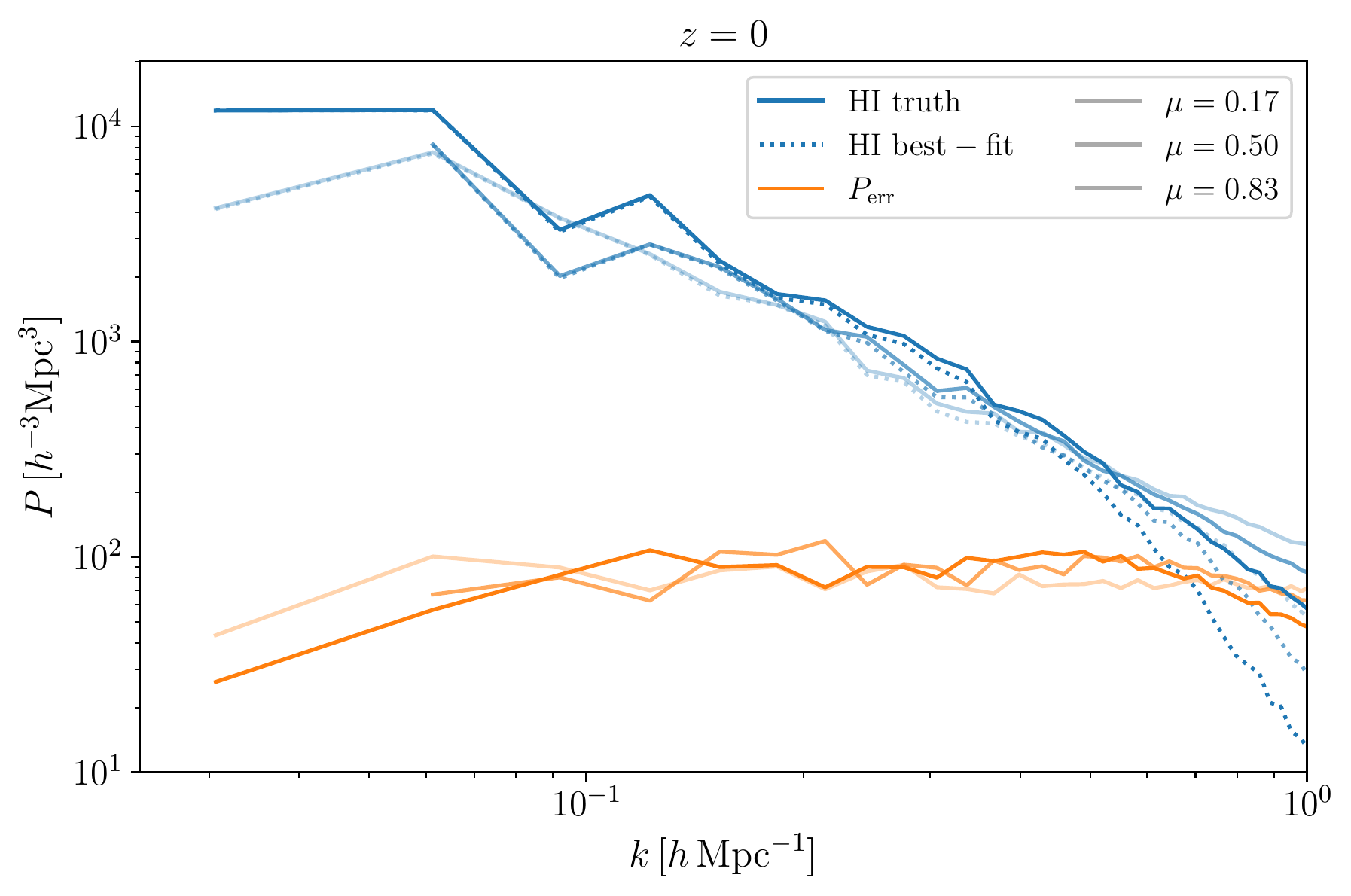}}
\subfloat{\includegraphics[width=0.48\textwidth]{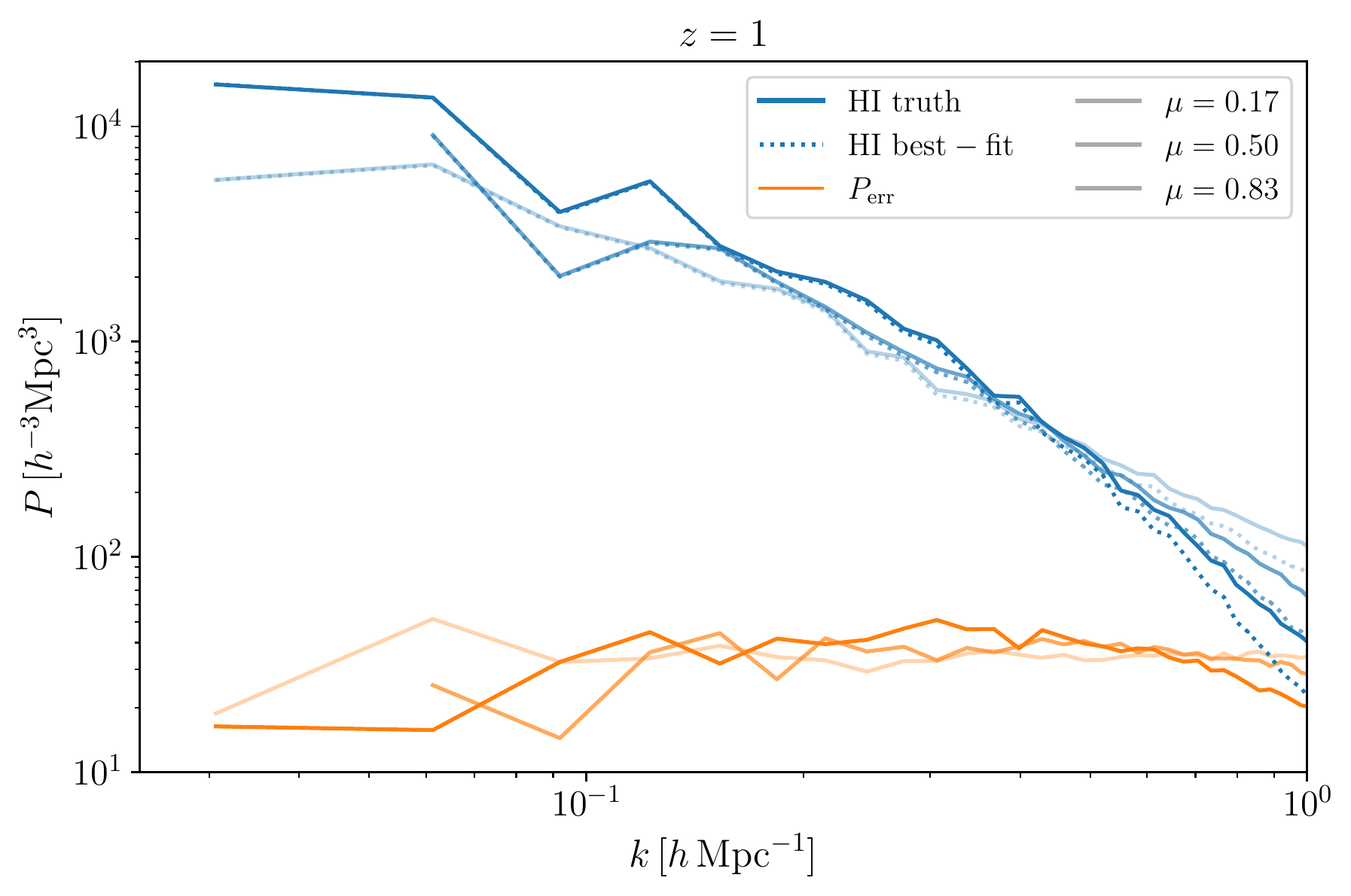}}
\caption{Redshift-space power spectrum: simulated, best-fit model and error power spectrum $P_\mathrm{err}(k,\mu)$ at $z=0$ (\textit{left panel}) and $z=1$ (\textit{right  panel}). The power spectra are measured in three wide~$\mu$ bins (different shading).}
\label{fig:rsd_pks}
\end{figure*}

In order to compare the anisotropic redshift space power spectrum to the theoretical prediction, we measure it in the three uniform wide~$\mu$ bins centered around $\mu=0.17$, $\mu=0.50$ and $\mu=0.83$. The results are shown in Fig.~\ref{fig:rsd_pks}. Even though the measurements have a large scatter, we find that perturbation theory predicts the measured power spectrum well on large scales. Looking at the error, we can see that it is still flat in $k$ on large scales, with similar amplitude (in all bins) as in real space, as expected. On small scales the deviation from a flat power spectrum is more prominent than in real space. This is a consequence of the nonlinear velocity which can produce large distortions in redshift space. We will comment more on this scale dependence in~\S\ref{sec:HInoise} and use it to estimate the nonlinear HI velocity dispersion. 

\begin{figure*}[!ht]
\subfloat{\includegraphics[width=0.96\textwidth]{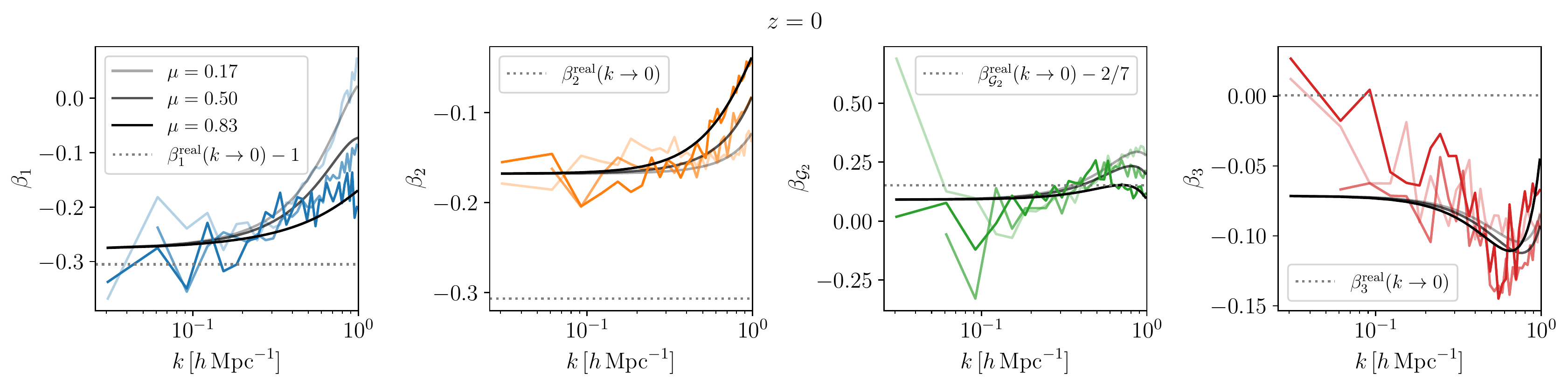}}\\
\subfloat{\includegraphics[width=0.96\textwidth]{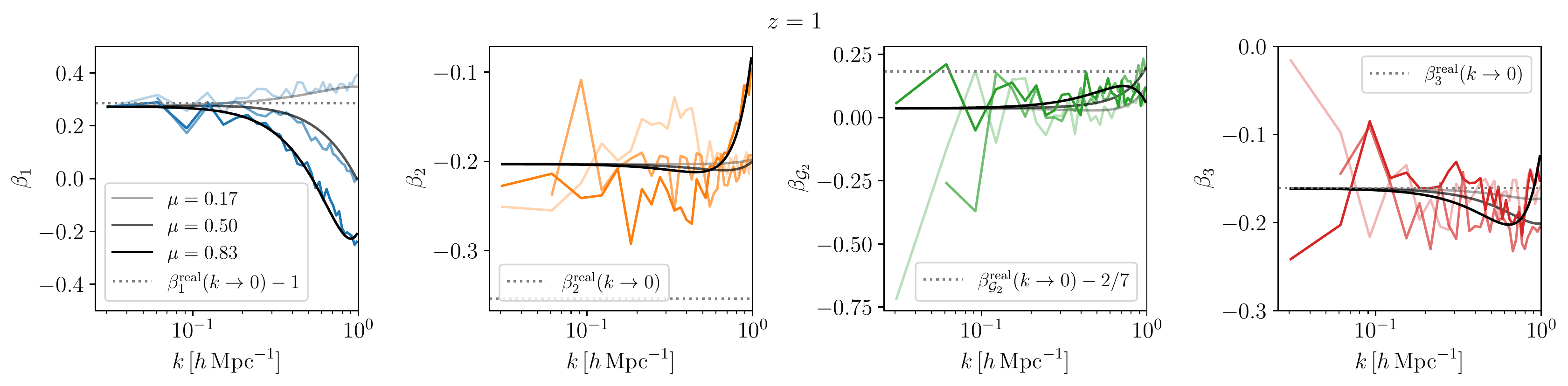}}
\caption{Redshift space transfer functions $\beta_i(k,\mu)$ for the cubic bias model at at $z=0$ (\textit{top}) and $z=1$ (\textit{bottom}) (lines with different colors). We show the best-fit transfer functions (black solid lines) using the following polynomials: $\beta_1(k,\mu) = a_0 + a_1k + a_2k^2 + a_{4}k^4 + a_{22}(k\mu)^2 + a_{44}(k\mu)^4$ and $\beta_i(k,\mu) = a_0 + a_2k^2 + a_{4}k^4 + a_{22}(k\mu)^2 + a_{44}(k\mu)^4$ for higher order transfer functions (see Table \ref{tab:polynomial_bestfits}). Different shading corresponds to different $\mu$-bins in all panels. For comparison we also show the low-$k$ values using polynomial fits in real space ($a_0$ values from Table \ref{tab:polynomial_bestfits}) (dotted lines).}
\label{fig:rsd_transfer}
\end{figure*}

In Fig.~\ref{fig:rsd_transfer} we show the redshift space transfer functions $\beta_i(k,\mu)$. We find that these transfer functions are rather smooth, both in $k$ and $\mu$. On large scales, $\beta_1$ and $\beta_{\G_2}$ have the same value in each $\mu$ bin, as a consequence of our choice to keep $\delta_Z^s$ and $\tilde{\mathcal{G}}_2^\parallel$ explicitly in the model for HI field in redshift space. As expected, $\beta_2$, $\beta_{\G_2}$ and $\beta_3$ are rather flat on large scales and they can be approximated as being constant. The first transfer function $\beta_1$ is more scale dependent as expected, but the measurements in different $\mu$ bins on large scales are very noisy, not allowing for a robust test of perturbative model in redshift space. For this reason, we will use a simple phenomenological fit for $\beta_1$ to account for the $k$ and $\mu$ dependence. We will come back to this and discuss details in~\S\ref{sec:mocks}. The bottom line is that, in the same way as in real space, the transfer function behaviour on large scales is compatible with perturbation theory expectations.

\begin{figure*}[!ht]
\subfloat{\includegraphics[width=0.48\textwidth]{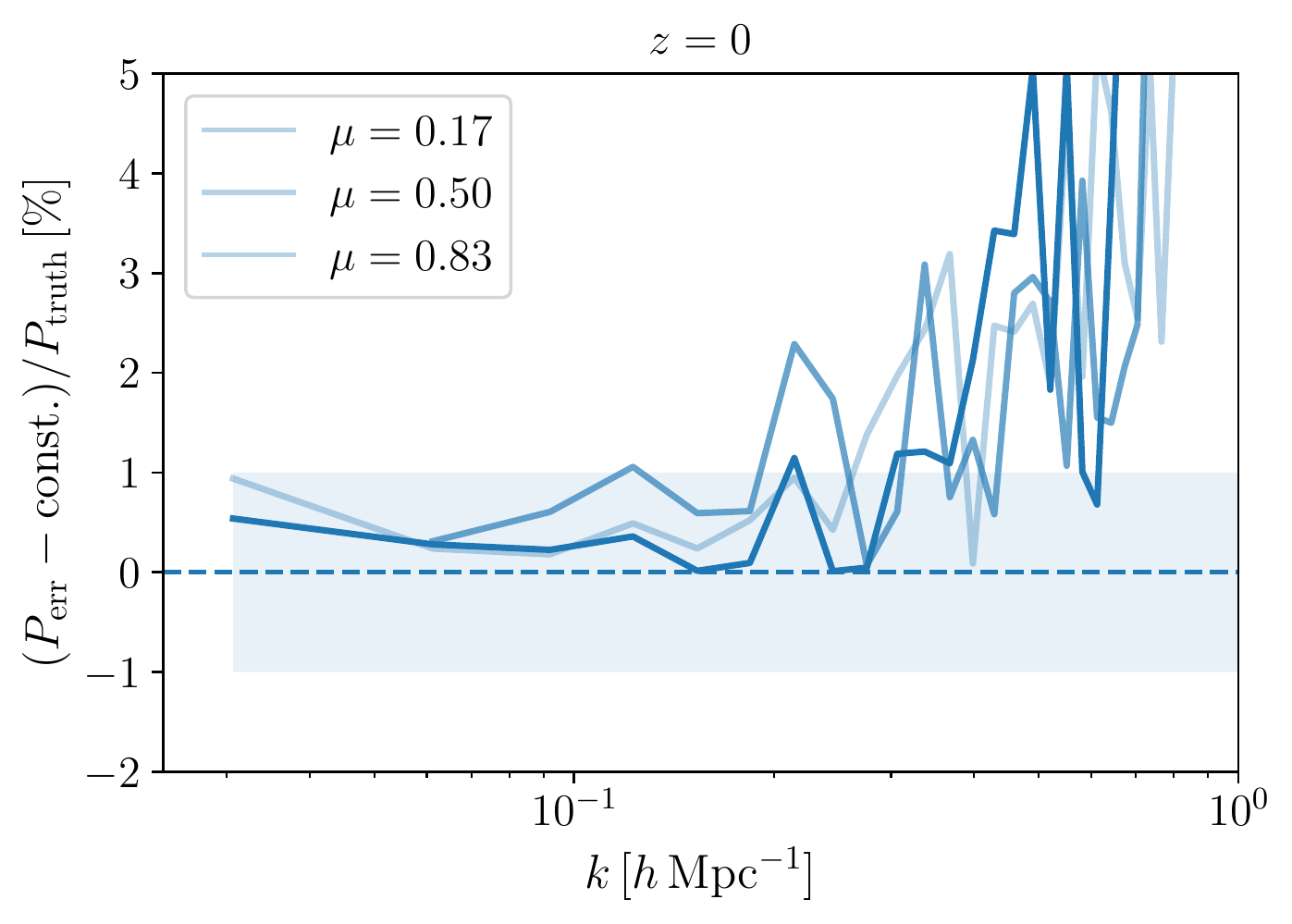}}
\subfloat{\includegraphics[width=0.48\textwidth]{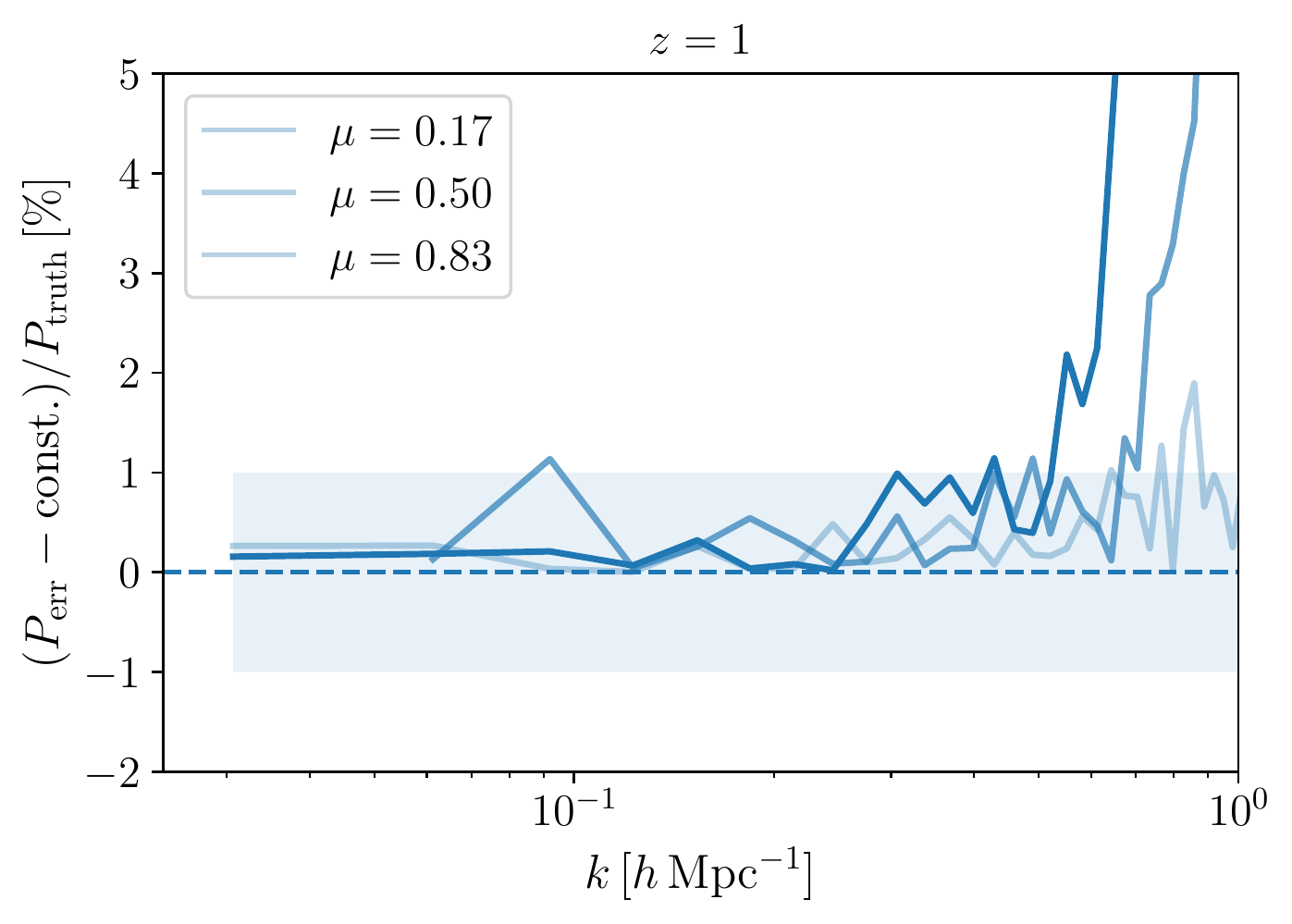}}
\caption{Fractional deviation from constant of the error power spectrum $\Perr$ using cubic bias model in redshift space with free transfer functions $\beta_i(k,\mu)$ at $z=0$ (\textit{left}) and $z=1$ (\textit{right}). The blue shaded region represents $\pm1\%$ of $P_\mathrm{HI}^\mathrm{truth}$.}
\label{fig:rsd_perr_const}
\end{figure*}

Similarly to what we did for real space, in Fig.~\ref{fig:rsd_perr_const} we show the relative error of the model after subtracting a constant $\Perr$ piece. The value of the constant is chosen as an average plateau of $\Perr(k,\mu)$ on $k=0.1-0.3\hMpc$ scales in each $\mu$-bin. As before, the fractional deviation of $\Perr(k,\mu)$ from constant allows us to infer at which scales we expect our PT model to break down. We can see that the cubic model is able to accurately model the true HI power spectrum within $1\%$ up to $k_\mathrm{max}\sim 0.3\ \&\ 0.4\hMpc$ at redshifts $z=0\ \&\ 1$, respectively. We can also see that the model is accurate up to smaller scales (higher-$k$s) for lower $\mu$ bins, as these modes are less affected by the nonlinear RSDs.

\begin{figure*}[!ht]
\subfloat{\includegraphics[width=0.48\textwidth]{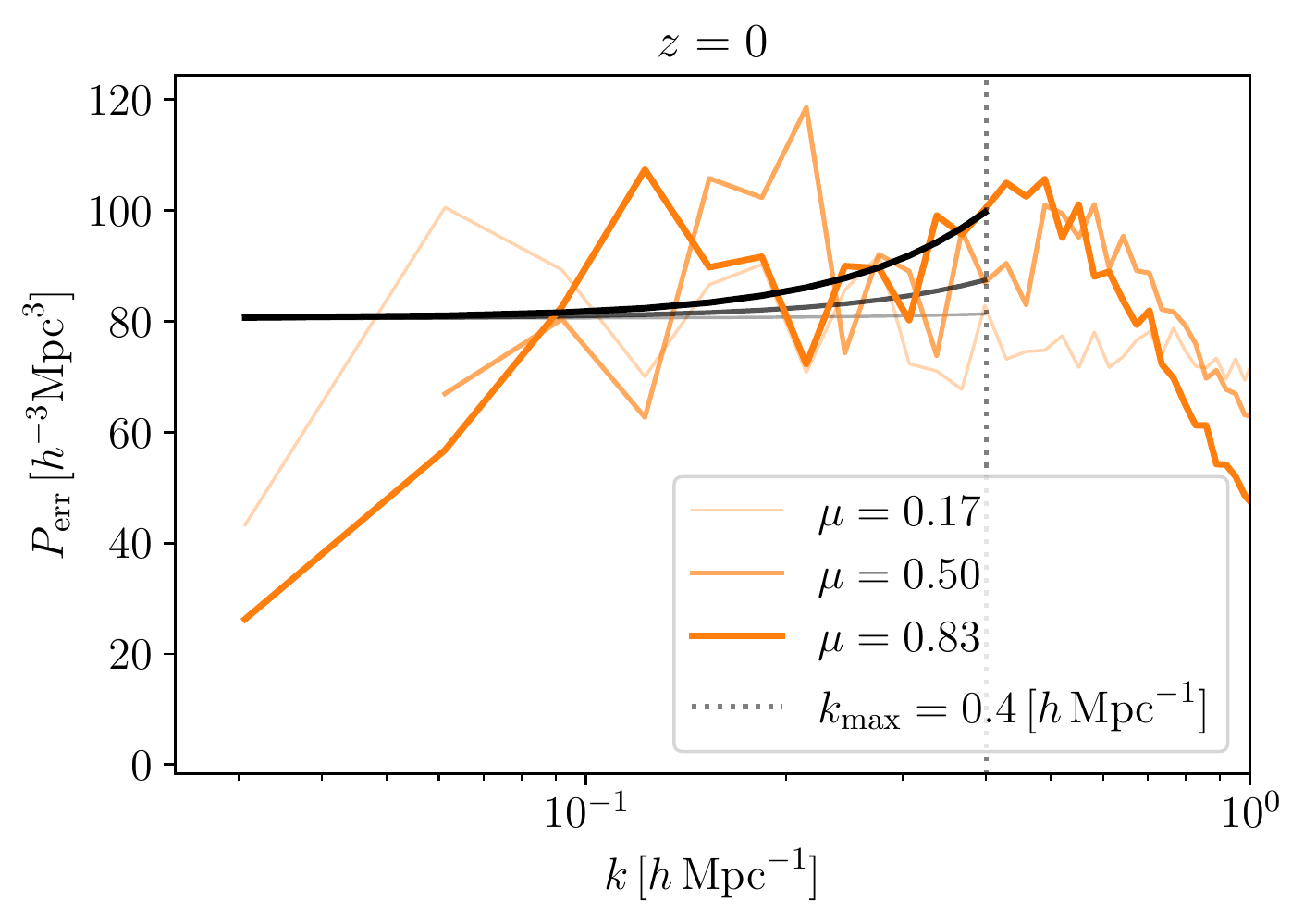}}
\subfloat{\includegraphics[width=0.48\textwidth]{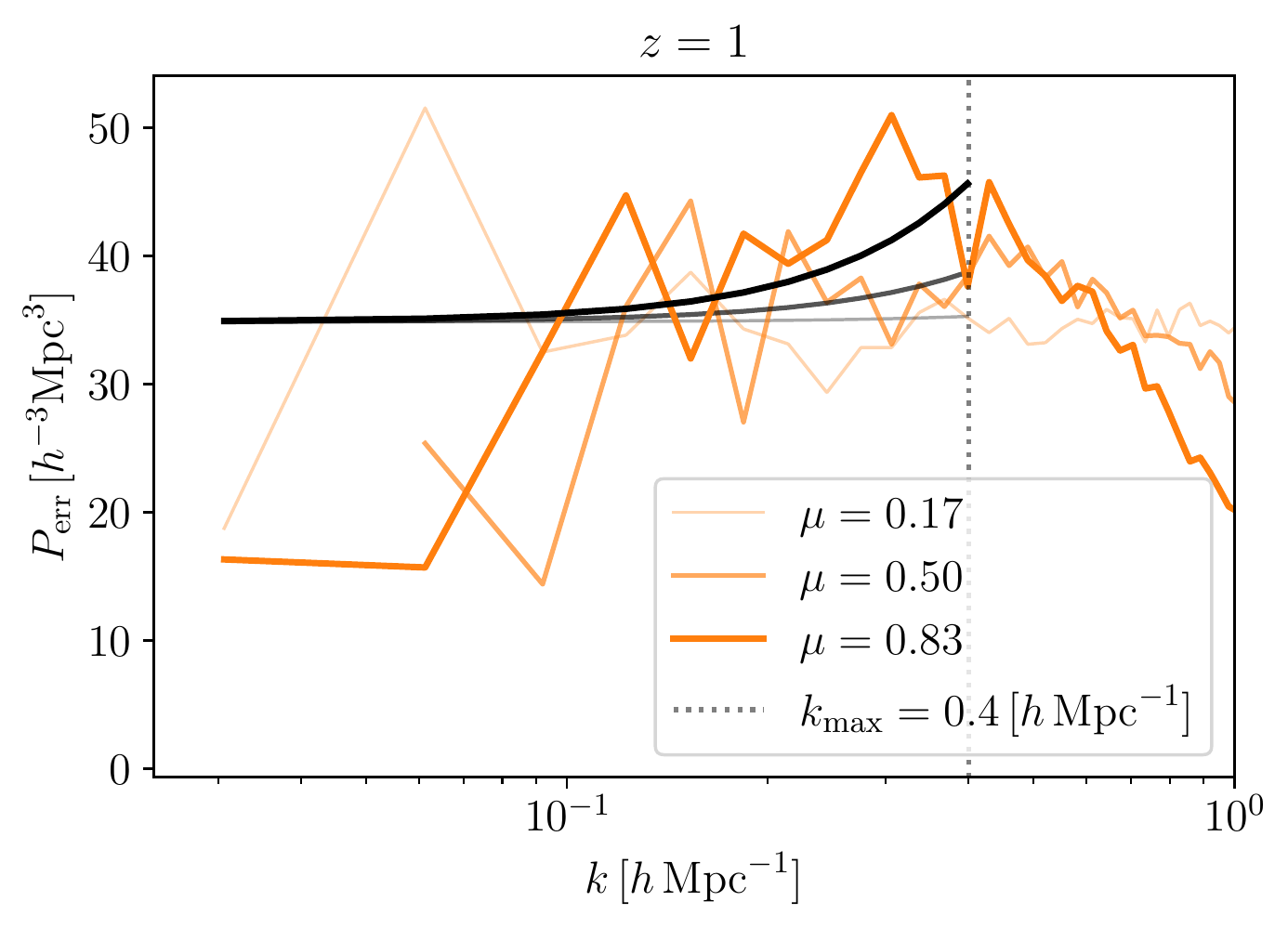}}
\caption{Redshift-space error power spectrum $P_\mathrm{err}(k,\mu)$ at $z=0$ (\textit{left panel}) and $z=1$ (\textit{right  panel}). The black solid lines show the best-fit model for $\Perr$ from Eq.~\eqref{eq:Perr_kmu} using $k_\mathrm{max}=0.4\hMpc$ (vertical dotted line). All power spectra are measured in 3 wide $\mu$ bins (orange lines) and different shadings and line widths correspond to different $\mu$ bins.}
\label{fig:rsd_perr_kmu2}
\end{figure*}

\section{HI noise properties}\label{sec:HInoise}

We have argued in the introduction that one of the main advantages comparing the theory and simulations at the field level is that such procedure allows to clearly isolate the model error. Understanding the properties of this error is equally important as having a good model, both for standard analyses based on $n$-point functions as well as for the more ambitious methods such as field-level forward modelling. In this section we study the properties of the error in detail and comment on its peculiarities in the case of the HI field. 

\subsection{Non-gaussianity of the error}\label{sec:NG_of_noise}

We can first have a look at the one point PDF of the model error. As we have seen in \S\ref{sec:1pt_pdfs} the cubic bias model provides a good description of the simulated HI field. However, we can also study the properties of the model error distribution. To obtain the model error distributions we compute the histogram of the residuals  ($\delta_\mathrm{HI}^\mathrm{truth}-\delta_\mathrm{HI}^\mathrm{model}$) for different 3D Gaussian smoothing scales and we show the results in bottom panels of Fig.~\ref{fig:hist}. We estimate the level of non-Gaussianity by comparing the model errors to the Gaussian distribution with matching variance and zero-mean. We find that using smaller smoothing scales the model error distribution deviates from the Gaussian distribution, especially in the tails. This means that assuming a Gaussian likelihood in the data analysis at the field level is not adequate. However, these deviations become smaller and the model errors become nearly Gaussian when larger smoothing scales ($R=5$ or $10\Mpch$) are used. Some discrepancies in the tails of the distributions are still present, although the number of discrepant pixels is relatively small.

\subsection{Amplitude of the noise on large scales}\label{sec:noise_amplitude}

In previous section we found the noise power spectrum to be flat in $k$. Here we want to discuss its amplitude in detail. Naively, one may think that the only fluctuations on large scales are the linear modes (multiplied by the linear bias) and random fluctuations due to the discreteness of the bias tracer. It is then natural to define the HI stochasticity as the power spectrum of the following difference: $\delta_{\rm HI}^{\rm truth} - b_1 \delta_{\rm m}$. Using the best fit value for $b_1$, this can be expressed in terms of the HI-matter cross-correlation coefficient $r_\mathrm{HI,m}$ \cite{Seljak+09,PacoTNG}:
\be
P_\mathrm{HI} - \frac{P^2_\mathrm{HI,m}}{P_m}= P_\mathrm{HI}(1-r^2_\mathrm{HI,m}),
\ee
where $P_\mathrm{HI,m}$ and $P_m$ are the HI-matter cross and matter power spectrum, respectively. Note that the stochasticity is equivalent to $\Perr$ for the linear bias model. 

\begin{figure*}
\subfloat{\includegraphics[width=0.48\textwidth]{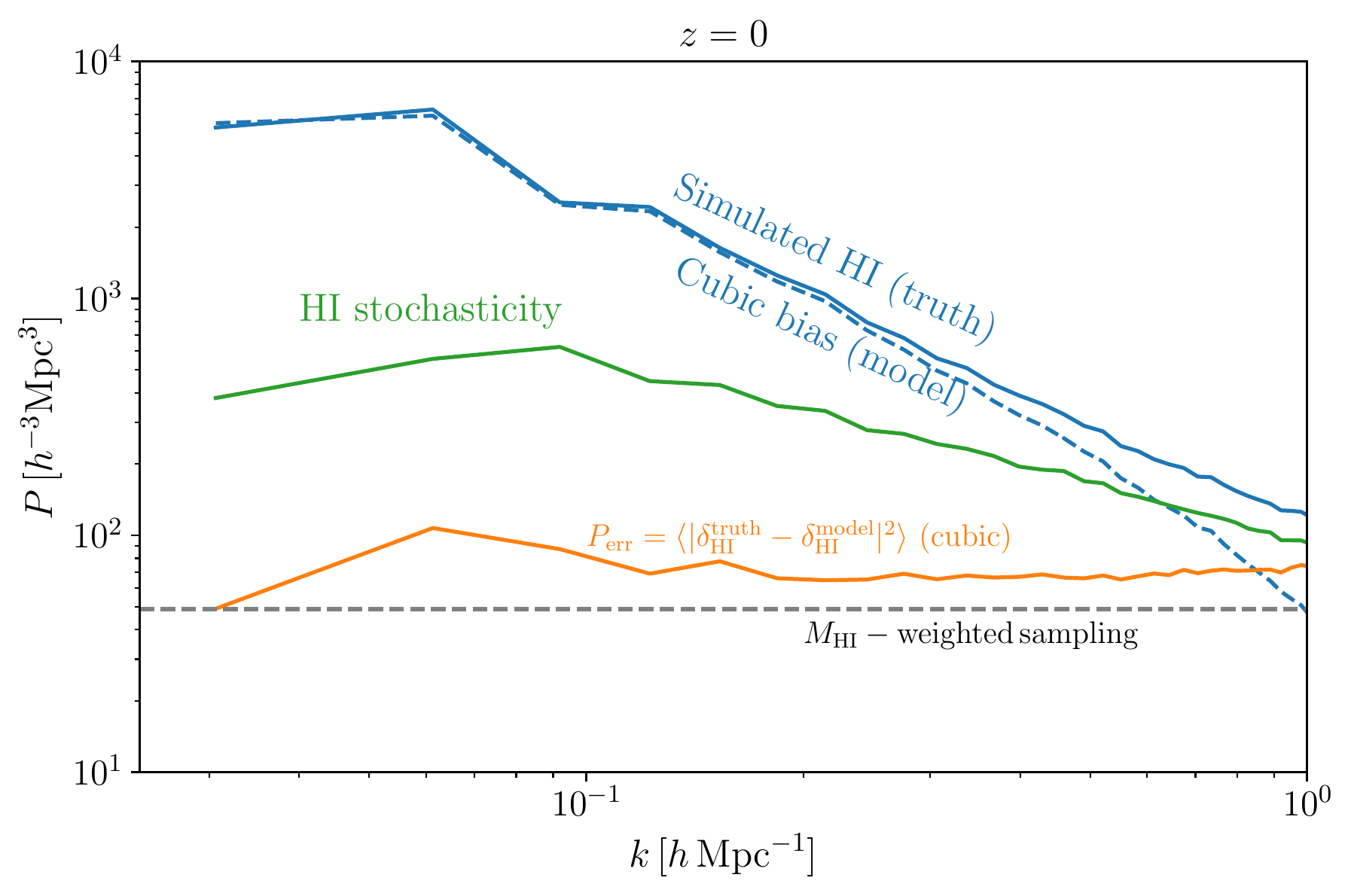}}
\subfloat{\includegraphics[width=0.48\textwidth]{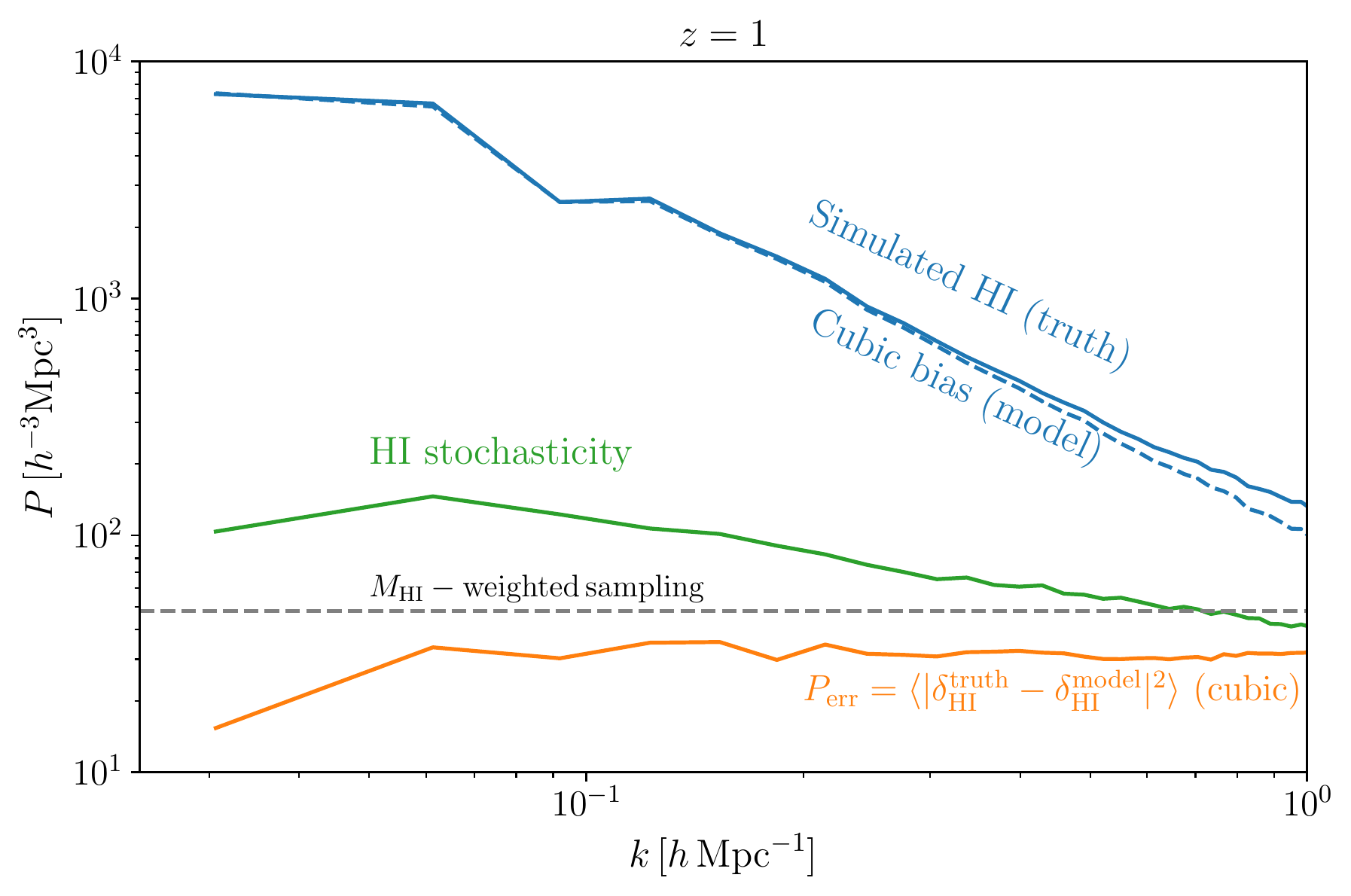}}
\caption{HI noise power spectrum estimates: $\Perr$ using the cubic bias model (orange), HI stochasticity $P_\mathrm{HI}(1-r_\mathrm{HI,m}^2)$ (green) and the $M_\mathrm{HI}$-weighted sampling noise (grey dashed). For comparison we show the simulated and best-fit $P_\mathrm{HI}$ (blue lines). We find $\Perr$ is almost an order of magnitude smaller than HI stochasticity on largest scales at both redshifts, while at $z=1$ $\Perr$ is smaller than the usually assumed HI noise power spectrum – $M_\mathrm{HI}$-weighted sampling noise.}
\label{fig:HI_noises}
\end{figure*}

However, the measured amplitude of the stochasticity is significantly larger than the naive sampling noise, given by $1/\bar n$, where $\bar n$ is the number density of $M_\mathrm{HI}$-weighted halos~\cite{PacoTNG}. We call this quantity $M_\mathrm{HI}$-weighted sampling noise and compute it following eq.~26 from Ref.~\cite{PacoTNG}. Furthermore we find good agreement between $M_\mathrm{HI}$-weighted sampling noise and the amplitude of high-$k$ plateau of $M_\mathrm{HI}$-weighted halo power spectrum, similar to what was found in the case of TNG100-1~\cite{PacoTNG}. The discrepancy between stochasticity and sampling noise is well-known for small-mass halos~\cite{2010PhRvD..82d3515H,Schmittfull,2017MNRAS.472.3959M,PacoTNG} and we show the measurements for Illustris in Fig.~\ref{fig:HI_noises}. This discrepancy between the $M_\mathrm{HI}$-weighted sampling noise and stochasticity clearly indicates that something is inconsistent in the simple picture we have just described. 

The resolution of this problem was given in~\cite{Schmittfull}. Through the nonlinear bias operators, such as $\delta_2$, the galaxy density field on very large scales gets contributions from small but perturbative modes. These long-wavelength fluctuations have the flat power spectrum which is indistinguishable from the sampling noise, but it can have much larger amplitude for very dense tracers.

To explicitly test this hypothesis, we can compare the error power spectrum in the model which includes quadratic biases to the $M_\mathrm{HI}$-weighted sampling noise. This is shown in Fig.~\ref{fig:HI_noises}. In the full quadratic and cubic model, $\Perr$ is indeed comparable to the Poisson noise. This means that at the map level, perturbation theory correctly predicts the modes which lead to large stochasticity. 

This has several important implications. First, the nonlinear bias is needed in order to have the correct map, not only on small scales in order to improve the model, but also on large scales in order to consistently predict all long-wavelength modes. Second, when calculating the covariance matrix for the power spectrum, the total noise-like contribution is much lager than the $M_\mathrm{HI}$-weighted sampling noise and indeed it is given by the stochasticity. While in simulations we were exploiting the fact that we knew the initial conditions, which allowed us to disentangle the contributions to the noise from the nonlinear bias and Poisson sampling, in a real analysis the initial conditions are unknown, and as a consequence the two contributions to the noise power spectrum are indistinguishable. This is an important observation, since it implies that the realistic values for the amplitude of the noise in the HI power spectrum Fisher forecasts is given by stochasticity and it can be an order of magnitude larger than what is usually assumed~\cite{highz21cm,PUMA}, making the forecasts overly optimistic. 

Finally, one can ask if an alternative analysis can be done such that the ``true'' noise, close to the $M_\mathrm{HI}$-weighted sampling noise, is the only price to pay in the total error budget. After all, since the long-wavelength fluctuations induced by the nonlinear bias are predictable, should not they be used as a signal, rather than treated as the noise? While this cannot be done in the power spectrum analysis, it is possible to achieve it including higher order $n$-point functions or doing the full forward modeling at the field level. As explained in~\cite{giovanni}, this situation is similar to the BAO reconstruction, where the information at the map level is used to sharpen the BAO peak, which can be measured in the 2-point function only with a limited precision.

\subsection{Nonlinear velocity dispersion estimates}\label{sec:velocity_dispersion}

As we already emphasised, the error on small scales can be dominated by the nonlinearities that are not taken into account in the model. The dominant nonlinearity in redshift space are Fingers of God induced by the nonlinear velocity dispersion $\sigma_{v,{\rm NL}}$. Their effect on the noise is to introduce a characteristic scale and angular dependence proportional at leading order to $\tfrac{1}{2} \sigma_{v,{\rm NL}}^2 f \mu^2 k^2$. We can use this fact to estimate the nonlinear velocity dispersion from the error power spectrum. We use the following model:
\be
\label{eq:Perr_kmu}
\Perr(k,\mu) = c_{\epsilon,1} \left( 1 + \frac 12  \sigma_{v,{\rm NL}}^2 f \mu^2 k^2 \right) ,
\ee
where $c_{\epsilon,1}$ and  $\sigma_{v,{\rm NL}}$ are free parameters. We obtain the best-fit values fitting $\Perr(k,\mu)$ and weighting each $k$-bin by $k$. We show the best-fit model of $\Perr$ in Fig.~\ref{fig:rsd_perr_kmu2}, while the best-fit values at both redshifts are presented in Table~\ref{tab:rsd_Perr_best_fit}.

\begin{table}[!ht]
    \centering
    \begin{tabular}{c|c|c}
    $z$ & 0 & 1\\
    \hline
    $c_{\epsilon,1}\ [h^{-3}\mathrm{Mpc}^3]$  & 80.6 & 34.8\\
    $\sigma_{v,{\rm NL}}\ [\Mpch]$ & 2.87 & 2.54\\
    \end{tabular}
    \caption{Best-fit parameters of the $\Perr(k,\mu)$ model at different redshifts obtained using $k_\mathrm{max}=0.4\hMpc$.}
    \label{tab:rsd_Perr_best_fit}
\end{table}

We find that the constant part of $\Perr$ is in agreement between real and redshift space. Our estimate of $\sigma_{v,{\rm NL}}$ shows that the HI nonlinear velocity dispersion is smaller than for the typical galaxy samples at similar~redshifts~\cite{Schmittfull_RSD,boss_ivanov}, in line with the conclusions of previous studies~\cite{PacoTNG}. However, this measurement has to be taken with a grain of salt, given the large numerical noise in the measured $\Perr(k,\mu)$ on large scales.

\section{HI mock data}\label{sec:mocks}

In all previous sections we focused on modeling based on perturbation theory. However, inspired by the perturbative templates (shifted operators), we can try to see how well such simple description works even on scales where formally perturbation theory does not apply. In other words, we can think of shifted operators as a phenomenological basis in which we can decompose the nonlinear field, with a set of smooth transfer functions. Even though unjustified in perturbative framework, such approach can be useful as a simple tool to easily generate realistic HI fields with the HI clustering properties similar to the ones of TNG simulation, even on small scales beyond $k_{\rm NL}$. In this section we show how to do this. 

For producing mocks, we will focus on scales up to $k_{\rm max}=1\hMpc$. The two main observations to keep in mind are that the measured transfer functions are smooth and they do not vanish at small scales, indicating that the HI field still correlates with the perturbative templates. Given the smoothness of the transfer functions, we can describe them using simple, low-order polynomial in $k$ and $\mu$. In the case of real space we use the following model
\begin{align}
\begin{split}
\beta_1^\mathrm{real}(k) &= a_0 + a_1k + a_2k^2 + a_4k^4,  \\
\beta_{i\ne1}^\mathrm{real}(k) &= a_0 + a_2k^2 + a_4k^4.
\end{split}
\end{align}
Note that in $\beta_1^\mathrm{real}$ we also include the term linear in $k$. This is needed since several terms in Eq.~\eqref{eq:beta1_model} have very mild scale dependence around $k\sim0.1\hMpc$, given the form of the loop corrections and the $\Lambda$CDM linear power spectrum. In the case of redshift space we include $\mu$-dependent terms equally for all transfer functions
\be
\beta_i^\mathrm{rsd}(k,\mu) = \beta_i^\mathrm{real}(k) + a_{22}(k\mu)^2 + a_{44}(k\mu)^4.
\ee

The polynomial coefficients $a_{i}$ and $a_{ii}$ are free parameters for each transfer function $\beta_i$, in both real and redshift space models. The total number of free parameters in the cubic bias model is thus 13 ($4+3\times3$) for real and 26 ($6+4\times5$) for redshift space, at a single redshift. We obtain the values of these parameters by doing the least-square fits to the measured transfer functions, weighting each $k$-bin by $k$. The polynomial fits for real and redshift space are shown in Figs. \ref{fig:best_fit_transfer_Perr} (black solid lines) and \ref{fig:rsd_transfer} (dashed lines), respectively. All the best-fit parameter values are shown in table \ref{tab:polynomial_bestfits}. We find that this simple model provides a good description of the measured transfer functions at both considered redshifts and for our choice of $k_\mathrm{max}=1\hMpc$.

\begin{table}[!ht]
    \centering
    \begin{tabular}{|c|c||c|c||c|c|}
    \hline
    \multicolumn{2}{|c||}{} & \multicolumn{2}{c||}{redshift space} & \multicolumn{2}{c||}{real space} \\
    \hline
    \multicolumn{2}{|c||}{$z$} & 0 & 1 & 0 & 1 \\
    \hline
    \multirow{5}{1em}{$\beta_1$} & $a_0$ & -0.28 & 0.27 & 0.69 & 1.28\\
    & $a_{1}$ & 0.09 & 0.09 & -1.04 & -0.52 \\ 
    & $a_{2}$ & 0.40 & 0.08 & 1.19 & 0.46 \\ 
    & $a_{4}$ & -0.17 & -0.04 & -0.28 & -0.02\\
    & $a_{22}$ & -0.53 & -2.11 & --- & --- \\
    & $a_{24}$ & 0.33 & 1.78 & --- & --- \\
    \hline
    \multirow{5}{1em}{$\beta_2$} & $a_0$ & -0.17 & -0.20 & -0.31 & -0.35\\
    & $a_{2}$ & -0.02 & -0.002 & 0.14 & -0.001 \\ 
    & $a_{4}$ & 0.02 & 0.01 & -0.03 & 0.03 \\
    & $a_{22}$ & 0.22 & -0.13 & --- & --- \\ 
    & $a_{44}$ & -0.13 & 0.45 & --- & --- \\
    \hline
    \multirow{5}{2em}{$\beta_{\mathcal{G}_2}$} & $a_0$ & 0.09 & 0.04 & 0.44 & 0.47 \\
    & $a_{2}$ & 0.54 & -0.09 & 0.05 & -0.38 \\ 
    & $a_{4}$ & -0.34 & 0.16 & -0.12 & 0.26\\
    & $a_{22}$ & -0.42 & 0.62 & --- & --- \\ 
    & $a_{44}$ & -0.20 & -1.02 & --- & --- \\
    \hline
   \multirow{5}{1em}{$\beta_3$} & $a_0$ & -0.07 & -0.16 & 0.001 & -0.16\\
    & $a_{2}$ & -0.12 & -0.02 & -0.06 &  0.24\\ 
    & $a_{4}$ & 0.11 & 0.02 & 0.03 & -0.14 \\
    & $a_{22}$ & -0.10 & -0.26 & --- & --- \\ 
    & $a_{44}$ & 0.23 & 0.48 & --- & --- \\
    \hline
\end{tabular}
    \caption{Best-fit parameters of the transfer functions when using the polynomial model in real and redshift space $\beta_i(k,\mu)$ at different redshifts obtained using $k_\mathrm{max}=1
    \hMpc$ (see Figs. \ref{fig:best_fit_transfer} and  \ref{fig:rsd_transfer}).}
    \label{tab:polynomial_bestfits}
\end{table}

Before we move on, let us comment on the inferred values of transfer function parameters in the low-$k$ limit. As we argued in \S\ref{sec:rsd_model_sec} we expect certain relations between real and redshift space transfer functions on large scales. This expectation turns out to be correct. As we can see in table~\ref{tab:polynomial_bestfits}, the following approximate relations hold
\begin{align}
\beta_1:\qquad a_0^{\rm real} &\approx a_0^{\rm rsd} +1 ,  \\
\beta_2:\qquad a_0^{\rm real} &\approx a_0^{\rm rsd} ,  \\
\beta_{\mathcal{G}_2}:\qquad a_0^{\rm real} &\approx a_0^{\rm rsd} + \frac 27 ,  \\
\beta_3:\qquad  a_0^{\rm real} &\approx a_0^{\rm rsd} ,
\end{align}
both at $z=0$ and $z=1$. Even though the agreement is not perfect, up to $\Delta a_0 \sim 0.1$, the real space and redshift space results are compatible, given the large numerical noise at low-$k$ in the measured transfer function. This is yet another confirmation of the consistency of the theory.

So far we have only considered the part of the HI signal that comes from the model prediction and correlates with the IC. However, we also need to include the stochastic part in order to generate a realistic HI mock. To do that we generate 3D fields with the power spectrum matching the inferred $\Perr$. In real space we generate a noise realization with an flat power spectrum amplitude of $\Perr=(68.53, 33.00)\ \,[h^{-3}\mathrm{Mpc}^3]$ at $z=(0,1)$, respectively. These amplitudes correspond to the $\Perr$ plateau between scales $k=0.1-0.3\ \hMpc$. In redshift space we generate a noise realization with the power spectrum constructed by doing a polynomial fit to the inferred $\Perr(k,\mu)$  (see Fig.~\ref{fig:rsd_perr_kmu2}). For the polynomial model we use the following ansatz
\be
\label{eq:rsd_perr_poly_fit}
\Perr(k,\mu) = a_0 + a_2k^2 + a_3k^3 + a_4k^4 + \sum_{i=2}^4a_{ii}(k\mu)^i,
\ee
 and perform a weighted least-squares fit up to $k_\mathrm{max}=1\hMpc$, weighting by $k$ each $k$-bin as before. The best-fit parameters values are shown in table~\ref{tab:rsd_Perr_poly_fit}. Note that this is different from the model that we used in Eq.~\eqref{eq:Perr_kmu} where the goal was to estimate HI velocity dispersion and where we used only the scales in the perturbative regime. With the polynomial model that we use here, the goal is to generate the most realistic noise in redshift space and be able to use it even on smaller scales. Finally, having generated the noise fields, we add them to the signal fields to obtain the final HI mock in either real or redshift space.

\begin{table}[!ht]
    \centering
    \begin{tabular}{c|c|c|c|c|c|c|c}
    $z$ & $a_0$ & $a_2$ & $a_3$ & $a_4$ & $a_{22}$ & $a_{33}$ & $a_{44}$\\
    \hline
    $0$ & 77.3 & 99.1 & -232.3 & 111.8 & 463.6 & -1282.1 & 861.9 \\
    $1$ & 36.6 & 16.4 & -50.8 & 29.1 & 106.4 & -317.6 & 206.7 \\
    \end{tabular}
    \caption{Best-fit parameters using the polynomial fit from Eq.~\eqref{eq:rsd_perr_poly_fit} to model $\Perr(k,\mu)$ at different redshifts.}
    \label{tab:rsd_Perr_poly_fit}
\end{table}

We can now use these fits to generate HI mocks. Since we measure the transfer functions without cosmic variance (using the same IC for the model and the simulated field), the transfer functions, and their smooth polynomial fits, can be applied to realizations with different IC as the dependence on IC is in the shifted fields. For any given realization of IC, we can then generate orthogonalized shifted fields and simply multiply them with the best-fit transfer function polynomials in order to obtain the HI signal map (cf.\ Eqs.~\ref{eq:real_model} and~\ref{eq:RSD_model}). The code we provide along this paper – \texttt{Hi-Fi mocks}, is performing exactly this procedure. 

This procedure has several free parameters. One is the grid size used to generate shifted fields and the other is the box size. In our fiducial analysis we use $N_\mathrm{mesh}=256^3$ which results in $\approx 16$ million particles. Given the size of TNG300-1 simulation ($L=205\ \Mpch$) this configuration makes generating HI mocks achievable even on personal computers in short amount of time and the full HI mock can be obtained on a modern laptop in few minutes. We note that for larger box sizes a larger grid size is needed in order to accurately probe smaller scales. This makes computation times longer and memory load larger.

We can contrast this procedure to the more standard way of generating HI mocks using N-body simulations and techniques such as halo occupation distribution, subhalo abundance matching and semi-analytical models. This is usually done by populating dark matter halos with HI mass after running the full N-body simulation. Approximate approaches to the full N-body simulations, such as COLA \cite{cola} or FastPM \cite{fastpm}, are able to accelerate this step. However the main bottleneck is running halo finder on top of simulation outputs which takes a comparable computational time. This step is not differentiable which makes using forward model in this setup unfeasible. Approaches based on machine learning techniques have been used to generate HI fields in cosmological boxes from dark matter field \cite{wadekar}, however these require training, testing and at the moment lack the interpretability. On the other hand our method is computationally cheap as it does not require running N-body, it does not require running halo finders, requires no training and in the low-$k$ limit it is based on perturbation theory. Furthermore it can be easily calibrated to other hydrodynamical simulations or simulations with different baryonic feedback prescription.

\section{Conclusions}\label{sec:conclusions}

In this paper we tested a perturbative forward model against the simulated HI at the field level. We found an excellent agreement on large scales, with sub-percent precision for $k<0.3\hMpc$ in real and redshift space, at redshifts of $z=0$ and $z=1$. This is confirmed looking at different statistics, such as cross correlation between the maps, the HI power spectrum and one-point probability distribution functions of the HI field. We also studied the properties of the HI noise. We confirmed the well-known fact that the stochasticity for HI can be much larger than the $M_\mathrm{HI}$-weighted sampling noise, and explained the origin of this discrepancy. We argued that this has an important implication for the HI power spectrum Fisher matrix forecasts and that it motivates the field level inference. Finally, using perturbation theory as an inspiration for the basis of templates at the field level, we find a phenomenological fit for the HI maps and provide a simple and efficient code to generate mock HI data for arbitrary initial conditions and volumes. Our results can be also seen as an important validation step for the HI modelling and they show that the currently existing pipelines for galaxy clustering analysis based on perturbation theory can be in principle applied to the neural hydrogen as well.

There are several interesting directions to explore in the future and we mention some of them here. First, we can use our \texttt{Hi-Fi mocks} code to generate realistic synthetic data in large volume boxes. The same code can be used to produce many realizations needed to estimate the covariance matrix. The currently existing pipelines for the power spectrum analysis can be then applied to these synthetic data. The resulting errors on the cosmological parameters would be the most realistic estimate of the constraining power of the HI power spectrum, even though in a very idealistic setup. A step towards a more realistic analysis requires inclusion of foregrounds, wedge and thermal noise, all of which can be added to the perturbative forward model. We leave this for future work.

Another direction is to explore the alternative summary statistics in more detail. We have already presented the results for one-point PDFs. Using predictions for the maps one can also study to what extent perturbation theory predicts the size function of voids, the nearest neighbour distribution etc. 

A natural question to ask is whether the perturbative forward model can be used to reconstruct the initial conditions. The simplest way would be to do perturbative inversion at the field level. This is along the lines of advanced reconstruction algorithms based on perturbative expansion, such as~\cite{2017PhRvD..96b3505S}. Alternatively, one can try to do the full froward modeling where the posterior for cosmological parameters is obtained by marginalizing over all amplitudes and phases of the initial conditions. The perturbative approach is useful in this context because it provides a simple differentiable forward model. In the future we plan to use our model to do the full field level inference.

It is worth mentioning that our analysis is based on a particular set of subgrid models used in the IllustrisTNG sumulation. It is known that other hydro-dynamical simulations show different baryonic effects \cite{baryonic_effects} on the matter field. Therefore, our results may change if other simulation or different astrophysical model is used. However, the effect of baryons is typically important on smaller scales, outside the validity of perturbation theory. Therefore, while the particular values of transfer functions will change when using different simulations, we do not expect this to limit the method or change our main results. In the future, we plan to repeat our analysis for simulations with larger volumes, more redshifts or different baryonic physics. This would lead to less noisy estimates of the transfer functions and provide an estimate of the impact of different galaxy formation scenarios.

\section*{Acknowledgments}
We thank Dylan Nelson and the IllustrisTNG team for providing initial random seed and linear power spectrum of the IllustrisTNG simulations, and for making the data publicly available through the online workspace\footnote{\href{https://www.tng-project.org/data/}{https://www.tng-project.org/data}}. We thank Francisco Villaescusa-Navarro for providing TNG100-1 HI density fields, TNG halo catalogs with $M_\mathrm{HI}$ masses and for the \texttt{Pylians}\footnote{\href{https://pylians3.readthedocs.io/en/master/}{https://pylians3.readthedocs.io/en/master/}} code library which we used to compute the HI density fields. We thank Mauro Bernardini, Giovanni Cabass, Simon Foreman, Uro$\mathrm{\check{s}}$ Seljak, Joachim Stadel, Matias Zaldarriaga for useful conversations. We thank Marcel Schmittfull for making the packages \texttt{perr}{\footnote{\href{https://github.com/mschmittfull/perr}{https://github.com/mschmittfull/perr}}}\cite{Schmittfull_RSD} and \texttt{lsstools}\footnote{\href{https://github.com/mschmittfull/lsstools}{https://github.com/mschmittfull/lsstools}} publicly available and acknowledge that these packages were particularly useful in our work. Furthermore, we acknowledge the use of \texttt{nbodykit} \cite{nbodykit}, \texttt{IPython} \cite{IPython}, \texttt{Matplotlib} \cite{Matplotlib}, \texttt{NumPy} \cite{Numpy,numpy2020} and \texttt{SciPy} \cite{SciPy}. This work made use of infrastructure services provided by $\mathrm{S^3IT}$ (www.s3it.uzh.ch), the Service and Support for Science IT team at the University of Zurich. AO acknowledges financial support from the Swiss National Science Foundation (grant no CRSII5{\_}193826). RF acknowledges financial support from the Swiss National Science Foundation (grant no 200021{\_}188552).

\appendix 

\section{Transfer functions in real space }
\label{app:real_app}
In this appendix we derive the form of the real-space transfer functions used in the main text. The starting point for our forward model is 
\begin{align}
\label{eq:app_def_HI_real}
\delta_{\rm HI}(\k) = \int d^3 \q\, (1+\delta_{\rm HI}^{\rm L}(\q))  e^{-i\k\cdot\left(\q+\vpsi(\q) \right) } \;,
\end{align}
where the HI field in Lagrangian coordinates~$\q$ is expressed as
\begin{align}
\label{eq:app_bias_expansion}
&\delta_{\rm HI}^{{\rm L}}(\q)\ =\ b_1^{{\rm L}}\,\delta_1(\q)\,+\,b_2^{{\rm L}}\,[\delta_2(\q)-\sigma_1^2]\, +\,  b_{\G_2}^{{\rm L}}\G_2(\q)\nonumber\\[4pt]
&\quad +\,b_3^{{\rm L}}\,\delta_3(\q)\,+\,  b_{\G_2\delta}^{{\rm L}}\,[{\G_2\delta}](\q)\, +\, b_{\G_3}^{{\rm L}}\, \G_3(\q)\,+\, b_{\Gamma_3}^{{\rm L}}\,\Gamma_3(\q)
\nonumber\\[4pt]
&\quad + b_{\nabla^2}^{\rm L} \nabla^2 \delta_1(\q)
\ .
\end{align}
The first line contains the linear and quadratic operators defined in the main text. Assuming the following form of bias operators
\begin{align}
\mathcal O (\k) = & \int_{\p_1,\p_2,\p_3} (2\pi)^3 \delta^D(\k-\p_1-\p_2-\p_3) \nonumber\\ 
& \quad F_{\mathcal{O}}^s(\p_1,\p_2,\p_3) \, \delta_1(\p_1) \delta_1(\p_2) \delta_1(\p_3) \;,
\end{align}
where $F_{\mathcal O}^s$ is a symmetrized kernel, the nontrivial cubic terms in the bias expansion have the following kernels
\begin{align}
F_{\mathcal G_3} &= \tfrac3{2} \tfrac{(\p_1\cdot\p_2)^2}{p_1^2p_2^2} - \tfrac{(\p_1\cdot\p_2)(\p_1\cdot\p_3)(\p_2\cdot\p_3)}{p_1^2p_2^2p_3^2} - \tfrac1{2} \; , \\
F_{\Gamma_3} &= \tfrac4{7} \left( 1 - \tfrac{(\p_1\cdot\p_2)^2}{p_1^2p_2^2} \right) \left( \tfrac{((\p_1+\p_2)\cdot\p_3)^2}{|\p_1+\p_2|^2p_3^2} -1 \right) \; .
\end{align}
Finally, the last term in Eq.~\eqref{eq:app_bias_expansion} is a higher derivative bias. Even thought it is linear in $\delta_1$, this term is suppressed by the derivatives. Note that it has the same form as the leading order counterterm for the dark matter field. 

The nonlinear displacement~$\vpsi(\q)$ can be expanded in perturbation theory too. The leading term is Zel'dovich displacement~$\vpsi_1(\q)$. Being the largest contribution it should be kept in the exponent, while higher order terms can be expanded. These higher order terms will mainly lead to operators which have the same form as in the bias expansion (but with fixed constants) and therefore can be absorbed in the bias coefficients. One exception is the second order shift acting on the HI density field, producing at cubic order
\be
\mathcal S_3 (\q) \equiv \vpsi_2(\q) \cdot \nabla \delta_1(\q) \;.
\ee
Such term cannot be written as a linear combination of bias operators and has the following kernel
\be
F_{\mathcal S_3} = - \tfrac{3}{14} \left( 1 - \tfrac{(\p_1\cdot\p_2)^2}{p_1^2p_2^2} \right)  \tfrac{(\p_1+\p_2)\cdot\p_3}{|\p_1+\p_2|^2} \;.
\ee

The uniform density displaced by Zel'dovich displacement in Eq.~\eqref{eq:app_def_HI_real} is just the Zel'dovich field, which can be written in perturbation theory as
\begin{align}
\delta_Z(\k) = \int d^3 \q\, \left( \delta_1 + \frac 12 \G_2 - \frac 13 \G_3  \right) e^{-i\k\cdot\left(\q+\vpsi_1 \right) } \;,
\end{align}
where all fields in the integrand depend on~$\q$. 

In conclusion, the real space model can be expressed in terms of the shifted bias operators and $\mathcal{S}_3$. The transfer function in Eq.~\eqref{eq:beta1_model} is then given by the sum of the linear bias, higher-derivative bias (and dark matter counterterm) and a set of all quadratic and cubic operators that correlate with~$\tilde\delta_1$. Note that those do not include $\tilde{\mathcal{G}_3}$, $\tilde{\mathcal G_2\delta}$ and $\tilde\delta_3$, since their correlation with $\tilde\delta_1$ either vanishes or renormalizes the linear bias. 

\section{Transfer functions in redshift space }
\label{app:redshift_app}
In this appendix we review the forward model in redshift space and give formulas for the transfer functions. In redshift space, the displacement in the exponent contains an additional term, with projected velocity along the line of sight
\begin{align}
\delta_{\rm HI}^s(\k,\mu) = \int d^3 \q\, (1+\delta_{\rm HI}^{\rm L}(\q))  e^{-i\k\cdot\left(\q+\vpsi(\q)+\frac{\hat \z\cdot \dot\vpsi(\q)}{H} \hat \z\right) } \;.
\end{align}
Using the same bias model in Lagrangian space as before and expanding all nonlinear displacement terms in the exponent, we get the following expression~\cite{Schmittfull_RSD}
\begin{align}
\label{eq:large_rsd_expression}
& \delta_{\rm HI}^s(\k,\hat\n) - \delta_Z^s(\k,\n) + \frac 37  f \tilde{\G}_2^{\parallel} (\k,\n) = \nonumber \\
& \int d^3 \q\, \Big[ \delta_{\rm HI}^{\rm L} -\frac 3{14} \G_2 - \frac 3{14} (1+b_1^{\rm L}) \delta_1\G_2 + \frac 16 \Gamma_3 + \frac 19 \G_3  \nonumber \\
& - \frac 37  f b_1^{\rm L} \delta_1 \G_2^{\parallel}  - \frac 58  f  \Gamma_3^{\parallel} + \frac 13  f  \G_3^{\parallel}  -  \frac 9{14} f \mathcal K_3  - \frac 3{14}  f^2  \delta_1^{\parallel}  \G_2^{\parallel}   \nonumber \\
& - R^{[2]}_{ij}\psi_2^i \partial_j \big( (1+b_1^L)\delta_1 + f\delta_1^{\parallel} \big)\Big]   e^{-i\k\cdot(\q+R^{[1]} \vpsi_1)}  \;,
\end{align}
where the operators projected along the line of sight are defined in the following way
\be
\mathcal O^{\parallel}(\q,\z) \equiv  \hat z^i \hat z^j \frac{\partial_i\partial_j}{\nabla^2} \mathcal O(\q) \;,
\ee
and the matrix $R^{[n]}$ is given by 
\be
R^{[n]}_{ij} (\hat\z) \equiv \delta_{ij} + nf\hat z_i \hat z_j \;.
\ee
We also define 
\be
\mathcal K_3(\q,\z) \equiv \hat z_i \hat z_j \frac{\partial_i\partial_m}{\nabla^2} \delta_1(\q) \frac{\partial_m\partial_j}{\nabla^2} \G_2(\q)  \;.
\ee
The first line in Eq.~\eqref{eq:large_rsd_expression} contains the terms that we model in the main text as a linear combination of shifted operators. The second line is the same as in real space. The third line contains new cubic operators with projection along the line of sight. Finally, the last term is the anisotropic second order shift which acts on the linear field. Note that for the new cubic operators in redshift space there are no new free parameters. This is consistent with the fact that velocities induced by gravity do not depend on the type of tracer. 

Using Eq.~\eqref{eq:large_rsd_expression} it is easy to read off the form of the transfer function $\beta_1^{\rm rsd}(k,\mu)$ in redshift space. It has the following form 
\be
\begin{split}
\beta_1^{\rm rsd}(k,\mu) = \beta_1^{\rm real}(k) + &\ \sum_i \frac{\vev{\shifted\delta_1 \shifted{\mathcal O}_i^\parallel }}{\langle\shifted \delta_1 \shifted \delta_1\rangle} - b_1 \frac{\vev{\shifted\delta_1 \shifted {\mathcal S}^{\rm rsd}_3}}{\vev{\shifted\delta_1 \shifted\delta_1 }}, 
\end{split}
\ee
where $\shifted{\mathcal O}_i^\parallel$ runs over all new cubic operators with projections along the line of sight, including the appropriate coefficients as given in Eq.~\eqref{eq:large_rsd_expression}, and $\shifted {\mathcal S}^{\rm rsd}_3$ is the operator in the last line of Eq.~\eqref{eq:large_rsd_expression}.

\section{Convergence test with respect to grid size}
\label{app:grid_convergence}

\begin{figure}[!ht]
\includegraphics[width=0.48\textwidth]{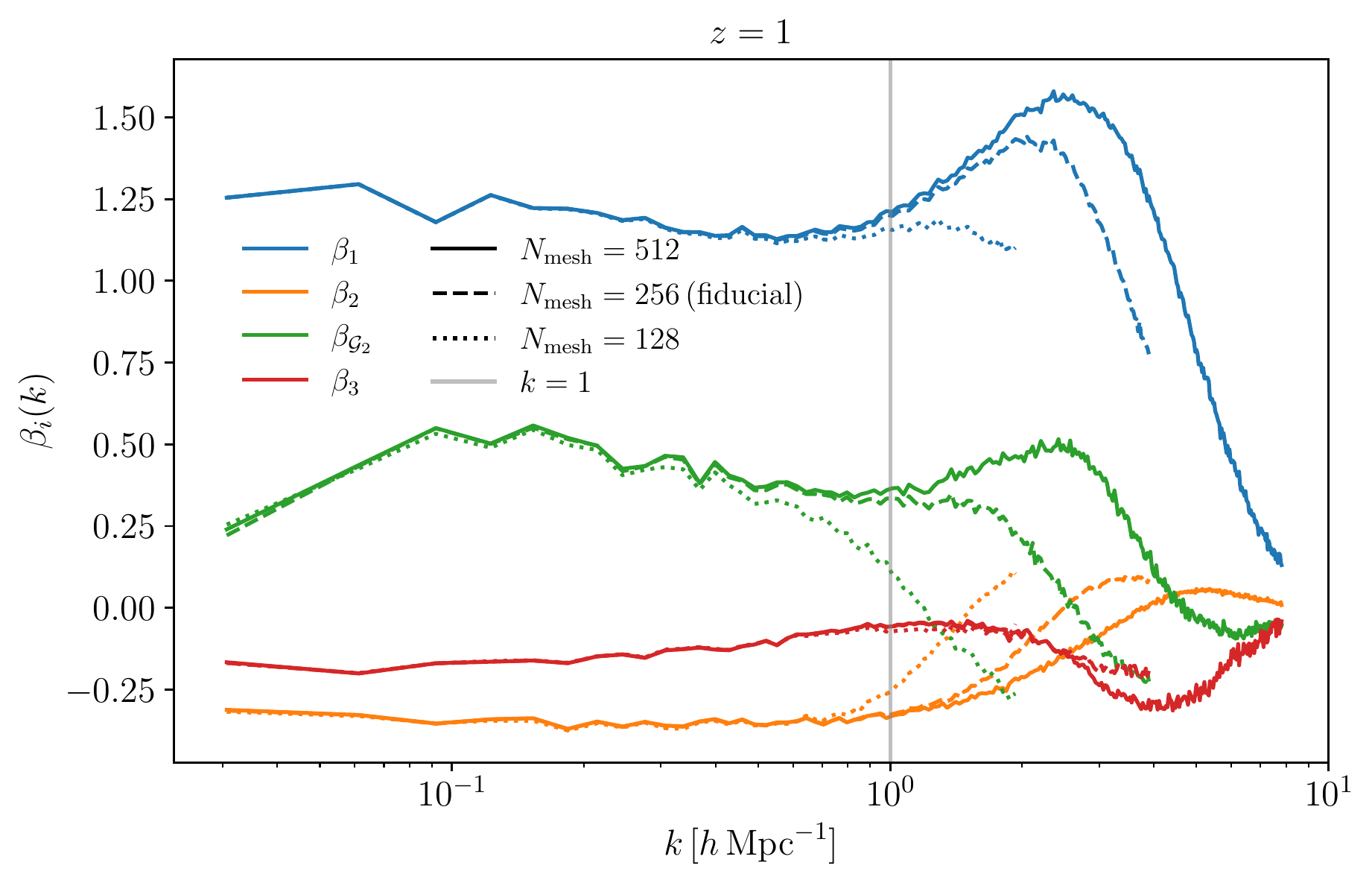}
\caption{Best-fit transfer functions $\beta_i(k)$ obtained from the simulated HI field at $z=1$ for different choices of grid size $N_\mathrm{mesh}$. We find that on relevant scales for this work $k < 1 \hMpc$ our results are converged with respect to the grid size for all transfer functions.}
\label{fig:convergence_test}
\end{figure}

In this appendix we demonstrate the convergence with respect to the grid size of our results on the scales we focus on. Throughout this work we use $N_\mathrm{mesh}=256$ grid cells per side (or equivalently the cell size of $0.8\Mpch$) to analyse simulated and predicted HI field. This choice determines the validity of our analysis. In order to test whether our results have converged we repeat our analysis using different number of grid cells per side. In particular we use lower ($N_\mathrm{mesh}=128$) and higher ($N_\mathrm{mesh}=512$) resolution and compare the resulting transfer function in Fig.~\ref{fig:convergence_test}, focusing on redshift $z=1$. As expected we find significant differences on very small scales and also when using a lower resolution grid. However, for the fiducial grid size and on scales $k < 1 \hMpc$ that we focus on in this work, we find our results to have converged.

\bibliography{References}
\bibliographystyle{JHEP}
\end{document}